%% file: DIP_1812_01_v2.tex
\documentclass[12pt,a4]{article}
\usepackage{amssymb}
\usepackage[dvips]{graphicx,color}
\usepackage{wrapfig}
\usepackage{makeidx}

\makeatletter
  \renewcommand{\theequation}{%
     \thesection.\arabic{equation}}
  \@addtoreset{equation}{section}
\makeatother

\input "butsuri_article_2018.tex"

\begin{document}

\bfr
DIP 1812-01\\
\efr

\begc
{\LARGE\bf  Restriction on Dirac's Conjecture}
\vs{0.3cm}

Takayuki Hori\footnote{email: hori@tokyo.zaq.jp}\vs{0.3cm}

Doyo-kai Institute of Physics, 3-46-4,
Kitanodai, \\
Hacniouji-shi, Tokyo 192-0913, Japan
\endc

 \vss

\begin{abstract}
First class constraints in a canonical
formalism  of a gauge theory might generate 
transformations which map a state to
its physically equivalent state.
This is called Dirac's conjecture.
There are two examples which may be candidates
of counter-example of the conjecture.
One is the toy model found by Cawley, and
another is the bilocal model proposed by
the author.
A quantum analysis of the bilocal model 
shows that the  model has the critical dimension of
spacetime, which is surprisingly equal to four. The derivation, however, is 
based on the assumption that true symmetry of
the system is generated by the first class
constraints, which holds if Dirac's conjecture is
satisfied.
In the present paper we give detailed and mathematically rigorous analysis of Dirac's conjecture in general
gauge theories, which involves new concept like
semi-gauge invariance.
We find the condition for the conjecture to hold.
This is a set of equations for the generating 
function of the transformation, expressed in terms of Poisson brackets and M-brackets introduced in the paper.
The above condition reduces the range of gauge theories 
where Dirac's conjecture holds.
Along with the general prescription described in the paper we find that the bilocal model  satisfies 
the above condition with some exceptions.
Some examples are used to illustrate our method.
\end{abstract}

\section{Introduction}

In 1964 Dirac conjectured that,
in the canonical theory of all gauge models,
every 1st class constraint may generate a transformation 
which maps 
a state into its physically equivalent 
one \cite{dirac_0,dirac_1}.
In those days no counter-examples were known,
though there has also been known no general proof of the conjecture. 
In 1980's there were some controversies among 
authors \cite{cawley_1,cawley_2,frenkel,sugano_1}, 
some of which claimed there is counter-example, while the others claimed the validity 
of the conjecture.
One of origins of the disagreement may be in the luck of
unique definition of hamiltonian, and
they discussed in such a way that one definition is more
{\it appropriate} than others. 
In a word the problem of Dirac's conjecture 
has not been defined in a mathematically rigorous 
manner.

In the present paper we 
establish the clear connection between transformations in
the lagrangian and the hamiltonian formalisms,
in general gauge theories,
based on the simplest definition of 
canonical hamiltonian.
There we use general solution of velocity variables
to the defining equation of canonical momenta.
Apart from the Poisson-bracket
we introduce the concept of M-bracket using Hessian matrices, which has informations on the degeneracy of a system having
gauge symmetries.
Dirac's conjecture is then examined in a mathematically
unambiguous way, and we give the conditions for the 
transformation in the phase space which maps a 
state into its physically equivalent one.
This makes us know what are the true physical 
symmetries of  gauge theories.

Before proceeding to the discussions of Dirac's
conjecture, it may be appropriate to 
explain the motivation for being interested in 
it, since all of the physically viable 
models at present may satisfy it.
In a bilocal particle model \cite{hori_1,hori_2,hori_3,hori_4,hori_5}
 proposed by the 
author, however, 
the lagrangian has two guage degrees of freedom,
while in the canonical theory there are
three 1st class constraints,
which may at first sight imply the breakdown of Dirac's
conjecture.
The phase space constraints generate  
$sl(2,\real)$ 
algebra 
which is the only one subalgebra of Virasolo
algebra of string model.
This fact may imply that 
the bilocal particle is only one
physically meaningful sub-entity of string.
An analisys of the quantume theory of the model 
indicates that it has 
critical dimension of spacetime being equal to four.
This is implied from 
the assumption that 
the physical symmetry of the model is 
$sl(2,\real)$.
The assumption is fulfilled if Dirac's conjecture
holds.
However, there exists at least one counter-example 
found by Cawley\cite{cawley_1,cawley_2},
which {\it does not} satisfy Dirac's conjecture.
What is the true physical symmetry 
of the bilocal particle?
 This is a critical question in the bilocal model to be answered,
though the problem of Dirac's conjecture 
for general gauge theories 
is interesting in its own right.
Fortunately, along with the general discussions 
presented here,
Dirac's conjecture in the bilocal model is shown to hold with some exception.

We restrict ourselves in the present papre
to the gauge models which have only
1st class constraints, and not have 
2nd class ones.
The 2nd class constraints are important
in a gauge theory, since the 1st class 
constraints become 2nd class after gauge
fixing.
There are super symmetric theories which 
have intrinsic 2nd class constraints,
where 1st class ones are not covariantly
seperated from 2nd class ones \cite{hori-kami,hori-kami-tate}.
In these theories the existence of 2nd class constraints gives rise  difficult problems
in the canonical theory.
We left the problem of 2nd class constraints
to future publications.

\section{Lagrangian theory}


Let us start with a general guage theory.
The action for the coordinate variables 
$q^{A},(A=1,2,..,N)$ 
and the velocity variables
$u^{A},(A=1,2,..,N)$
is 
\be
I
&=&
\int d\t L(q,u).
\ee
The Euler-Lagrange equations (ELE) are
\be
[{\rm EL}]_{A}
\deff
\fr{d}{d\t}\fr{\del L}{\del u^{A}}
-
\fr{\del L}{\del q^{A}}
=
M_{AB}\dot{u}^{B}
+
\w_{A}
=
0,
\eq{1_ELeq}
\ee
with
\be
M_{AB}
\deff
\pdel{W_A}{u^B}
,
\si
\w_{A}
\deff
\pdel{W_A}{q^B}
u^{B}
-
\pdel{L}{q^A}
,
\si
W_A 
\deff
\pdel{L}{u^A}
,
\eq{1_def_M_w_W}
\ee
where 
dots denote derivatives with respect to time,
$\t$,
and the repeated indices stand for
summations over the indices, $A=1,..,N$.
The matrix $M_{AB}(q,u)$ is called Hessian.
(In a field theory 
the indices include the spacial coordinates,
the derivatives stand for the 
functional derivatives and 
the summations are  integrals over
spacial coordinates.)
It is possible to consider the lagrangian as
a function of $(q,\dot{q})$,
but the discussions below become more clear
if the ELE's are
1st order differential equations of
$(q,u)$ with respect to $\t$.
The base space is spanned by independent 
variables $(q,u)$, and we denote ${\rm [EL]}_A=0$
together with $\dot{q}^A - u^A=0$
as ${\rm [EM]}=0$.
The classical
orbit is represented by a solution to 
the equations ${\rm [EM]}=0$.

In a gauge theory the Hessian is not 
a regular matrix, and 
the initial value problem of the 
ELE's has not 
always solutions for arbitrary initial 
values of $(q,u)$.
The system of linear algebraic equations
$\bbM\dot{\bbu}+\bbomega=\bbzero$
has solutions for $\dot{u}$'s
if and only if
$
R
\deff
{\rm rank}~\bbM
=
{\rm rank}~(\bbM,\bbomega).
$
Not only the 
initial values but the values at any point 
on the solution orbit 
must satisfy the above condition in order that the system 
has solutions, otherwise the ELE's 
contain contradiction.
This is the problem of integrability 
of the ELE's,
and was extensively discussed in ref.\cite{sugano_1}.
The authors of ref.\cite{sugano_1} obtained
the conditions for the suitable time development operator which is compatible with
the constraints, in a step by step method.

In the present paper,
we give a simple closed expression of 
the conditions of the initial values, which must hold for the integrability.
First note that the general solution for 
$\dot{\bbu}$'s to the equation
$\bbM\dot{\bbu}+\bbomega=\bbzero$
is 
obtained by the standard sweep out method.
We can find the regular matrix $\bbQ$, whose 
coefficients are functions of $(q,u)$, 
and by which the Hessian transformes to the form
\newfont{\bbbg}{cmr10 scaled\magstep2}
\newcommand{\bbzerol}{\smash{\hbox{\bbbg 0}}}
\newcommand{\bbzerou}{\smash{\lower1.7ex\hbox{\bbbg 0}}}
\newcommand{\bbem}{\smash{\hbox{\bbN}}}
\def\bbLamda{\bm{$\L$}}
\def\rkakkob{\left. \begin{array}{c}}
\def\rkakkoe{\end{array}  \right\} }
\be
\overbrace{\si\sii\sii\si~}^{R}~~
\overbrace{\sii\si~}^{N-R}
\sii\sii\si\si
\nn
\bbQ\bbM\bbC
=
\left(
       \begin{array}{c|cc}
         \begin{array}{cccc}
             1 & & & \bbzerou \\
               & 1 & &         \\ 
                   & & \ddots &    \\ 
             \bbzerol  &&& 1  
         \end{array}  
         &&  ~~  \bbem ~~ \\ \cline{1-3}
         &&\\ 
         \bbzerol &&  \bbzerol \\
       \end{array}
\right)
\!\!\!\!
\begin{array}{cc}
\rkakkob \\ \\ \\ \\\rkakkoe & \!\!{}_R\si~~\\
\rkakkob \\  \\\rkakkoe & \!\!\!\!\!\!{}_{N-R}
\end{array},
\eq{1_newHess}
\ee
where 
$\bbC$ is a constant matrix 
which may interchange columns of the matrices 
to which $\bbC$ acts from right.
Since $\bbC$ can be set to unity by properly arranging 
the order of the variables $(q,u)$, we set $\bbC=\bb1$ 
in what follows.
Then the general solution to
$\bbM\dot{\bbu}+\bbomega=\bbzero$
is the sum of a special solution and
linear combination of the solution to
the homogeneous equations
$\bbM\dot{\bbu}=\bbzero$.
Thus we have
\be
\dot{u}^a &=& -N^{a}{}_{m}v^m
- Q^{aB}\w_B,
\sii (1\le a\le R)
\eq{2_dotu^a=}
\\
\dot{u}^m &=& v^m(q,u),
\sii\sii (R+1\le m\le N)
\eq{2_dotu^m=}
\ee
where $v^m$ are arbitrary functions of $(q,u)$.
Here and hereafter we use the rule that repeated indices 
of first alphabets $a,b,..$ are summed over $1\sim R$, and those of later alphabets 
$m,n,..$ are summed over $R+1\sim N$.
(\ref{2_dotu^a=})
and
(\ref{2_dotu^m=})
are solution to the algebraic equations,
$\bbM\dot{\bbu}+\bbomega=\bbzero$,
if and only if
$
{\rm rank}~\bbM
=
{\rm rank}~(\bbM,\bbomega),
$
which means $Q^{mA}\w_A=0,(R+1\le m \le N)$.
These constraints are expressed in terms of
the eigen vectors of the Hessian with zero
eigen value,
$\bbz^{(m)},~(m=R+1,..,N)$, as
\be
\bbz^{(m)}
\cdot
\bbom
=
0,
\sii
(R+1\le m\le N),
\eq{2_LagCon1}
\ee
where the components of $\bbz^{(m)}$
are defined by
\be
z^{(m)A}
\deff
Q^{mA},
\sii
(m=R+1,..,N).
\eq{1_def_z}
\ee
The vectors $\bbz^{(m)},~(m=R+1,..,N)$ 
are linearly independent because 
$\bbQ$ is regular.

Since the conditions (\ref{2_LagCon1}) 
must be satisfied at all time,
the arbitrary order of derivatives 
of $\bbz^{(m)}\cdot\bbom$
with respect to $\t$ must vanish.
These are written as
\be
&&
D^{k-1}(\bbz^{(m)}\cdot\bbom) = 0,
\sii (m=R+1,..,N;~k=1,2,..)
\eq{2_2ndLagCon}
\\
&&
D \deff
u^{A}
\fr{\del}{\del q^{A}}
-
\left(
Q^{aB}\w_{B} + v^{n}N^{a}{}_{n}
\right)
\fr{\del}{\del u^a}
+
v^{n}
\fr{\del}{\del u^n},
\sii
\eq{2_def_D}
\ee
where we use (\ref{2_dotu^m=}).
The conditions (\ref{2_2ndLagCon})
are classified in three cases, where they are
(1) identities,
(2) conditions for the arbitrary functions $v^m$,
(3) new constraints for $(q,u)$.
We do not consider the case (2),
because it occurs when there are 
2nd class constraints in the canonical
formalism.
In the case (3), 
denote the independent equations 
among (\ref{2_2ndLagCon})
as
\be
\ell^{(k)}_{i}(q,u) = 0,
\sii (i=1,..,N_k;~k=1,2,...)
\eq{1_LagCon}
\ee
and call them 
$k$th order {\it lagrangian constraints}
(LC) \cite{kami_1}.

If (\ref{1_LagCon}) are satisfied for the initial values  of $(q,u)$,
then they are satisfied at all points 
on the solution orbit
as is seen by their construction.
Thus the necessary and sufficient condition
for the initial value problem of the 
ELE's to have solutions 
is that all of the $k$th order 
LC's are satisfied 
for the initial values of $(q,u)$.


Now let us consider transformations of 
$(q,u)$ with arbitrary infinitesimal parameters,
$\e$'s, containing the velocity variables:
\be
\d q^{A}
=
\e^{A}(q,u),
\sii
\d u^{A}
=
\fr{d}{d\t}
(\d q^{A}),
\eq{2_LagTR}
\ee
which keeps the relations $\dot{q}^A = u^A$.
This map is, in general, not a
transformation in the velocity-coordinate
space because it contains $\dot{u}$'s 
through $\d u$'s.
However,
this type of transformations 
is frequently considered in a wide class of
gauge models because there
are cases where the transformed  
lagrangian contains $\dot{u}$ 
dependent terms only as 
$\t$-derivative of some function,
which have no effect on the action.
It is easily shown that
the necessary and sufficient condition
for the existence of such a function is
\be
M_{AC}\pdel{\e^C}{u^B}
=
M_{BC}\pdel{\e^C}{u^A}
.
\eq{2_integCond}
\ee
In fact 
(\ref{2_integCond})
is the integrability condition 
for the existence of the function $E$
such that 
\be
W_{A}
\fr{\del \e^{A}}{\del u^{B}}
=
\fr{\del E}{\del u^{B}}.
\eq{1_CondLTR}
\ee
The variation of lagrangian has the form
\be
\d L
\deff
\D L + \fr{dE}{d\t},
\ee
where 
\be
\D L
=
\fr{\del L}{\del q^{A}}\e^{A}
+
\left(
W_{A}
\fr{\del\e^{A}}{\del q^{B}}
-
\fr{\del E}{\del q^{B}}
\right)u^{B}
\eq{2_defDL}
\ee
does not depend on $\dot{u}$.
The transformations defined by 
(\ref{2_LagTR})
with infinitesimal parameters 
satisfying (\ref{2_integCond}) 
are called 
{\it lagrangian transformations} \cite{kami_1},
or LTR for short.
Let us call $E$ the asocciated function of 
$\e^{a}$, which is determined up to arbitrary 
additive functions of $q$'s.

For example
the transformation defined by
\be
\d q^{A} = \ve_mz^{(m)A},
\sii
\d u^{A} = \fr{d}{d\t}\d q^{A}
\eq{2_primarySGTR}
\ee
is a LTR,
since (\ref{2_integCond})
holds because $\bbz^{(m)}$
are eigen vectors of the Hessian with
zero eigen value.
The variation of the lagrangian under the
transformation is
\be
\D L = -\ve_m\bbz^{(m)}\cdot\bbomega.
\eq{2_DL_primaryLTR}
\ee
Note r.h.s. of (\ref{2_DL_primaryLTR}) 
vanishes if LC's hold.
Let us call, in general, the LTR
under which 
$\D L$ vanishes up to 
LC's as {\it semi-gauge transformation},
or SGTR for short.
If $\D L=0$
the transformation is called {\it gauge}
transformation, or GTR for short.
The concept of the SGTR
plays an essential role in 
determining the physically equivalent classes
of states in a gauge theory as is shown below.

Now let us consider relation between the 
physical states and variables describing them.
Since the lagrangian dynamics determines
the time development of state,
the velocity variables $u^m,(R+1\le m\le N)$
whose time developments are arbitrary 
can not be used to describe any states.
Therefore we assume that they are unphysical
variables.
The corresponding coordinate variables
$q^m,(R+1\le m\le N)$
should also be unphysical.
Hence the two sets of variables $(q,u)$
and $(q',u')$ are called 
{\it physically equivalent} 
if they are different only by the
$m$-th components, $R+1\le m \le N$,
and write as
\be
(q,u)^a = (q',u')^a \si (1\le a \le R)
~\Longleftrightarrow
(q,u) \stackrel{\rm P}{\sim} (q,u').
\eq{2_PEinLag}
\ee
The values of the unphysical variables
are set freely, and the time development
of them are described by the free 
parameters $v^m$ in
(\ref{2_dotu^m=}).

A subtle but an important point should be
noted here.
There are cases in which 
LC's determine
the values of the unphysical variables 
in terms of the physical ones.
However,
as is shown at the last of the next section,
the  unphysical variables
have the gauge degrees of freedom,
and we can set arbitrary values for
the unphysical variables though 
the values of physical variables change
according to the GTR.
The latter may be written by the same symbols.
In a word the values of the unphysical variables 
can be set freely.
This is not the case in the model 
which have 2nd class constraints.

For example,
$A^0$ in the Maxwell model 
is unphysical variable since
ELE's do not determine
the time development of
$u^0=\dot{A}^0$.
The LC is the Gauss law 
which is a Poisson equation for $A^0$,
and completely determines the value of $A^0$ 
if one sets a boundary condition for $A^0$
at infinity.
The initial value problem in this case is
determined by the ELE's of the physical variables $\bbA,~\bbu$
and the equations 
$A^0 = c,~u^0 = \dot{c}$ with arbitrary
function $c$, which are
subjected to the LC.
The state described by $\bbA,A^0$
is physically equivalent to the state
described by $\bbA + \nabla\L,~A^0 + \del^0\L + \h$,
with arbitray $\L$ and $\h$,
since $\h$ is regarded as the variation of $c$.

According to Dirac
the physical equivalence
is extended in such a way that
two variables are physically equivalent
if they describe points on the 
solution orbits of ELE's, which have 
physically equivalent initial values
in the sense of
(\ref{2_PEinLag}).

In the present paper we explore
the transformations
by which a state maps to its
physically equivalent state.
Let us call such transformation
{\it physically equivalent transformation},
or PETR for short.
The simplest example of PETR is
of the form
\be
&&
\d_{\rm L}q^a = 0,
\sii
\d_{\rm L}q^m = \e^m(q),
\\
&&
\d_{\rm L}u^A = \fr{d}{d\t}\d_{\rm L}q^A.
\si (a:1\sim R,~m: R+1 \sim N)
\sii\si
\ee
This is a LTR because $\e^m$ do not
depend on $u$,
and is a PETR because
it moves only the unphysical 
components $q^m,~u^m,~(R+1 \le m \le N)$.
Another example of PETR is the gauge 
transformation.
This fact is proven as follows.
Let us denote O$_1$ the solution orbit
of ELE, O$_2$ 
the gauge transform of O$_1$
and O$'_2$ arbitrary transform of O$_2$.
Assuming three orbits start from the same point
and end at another point,
the values of the action, $I_1,I_2$ and $I'_2$, calculated along the orbits O$_1$,O$_2$ and O$'_2$,
satisfy
$
I'_2 = I_1 + {\rm O}(\e^2,\e_{\rm g}^2),
~
I_2 = I_1 + {\rm O}(\e^2_{\rm g}),
$
where $\e$ is parameter of arbitray transformation and
$\e_{\rm G}$ is that of gauge transformation.
Hence
$
I_2 = I'_2 + {\rm O}(\e^2),
$
so
the action has stationary value on
O$_2$,
which means O$_2$ to be a solution orbit
to ELE.
Thus we see a gauge transformation is a
PETR.

In many examples the set of PETR is wider than
that of GTR.
As is shown in the next section,
transformations generated by the 1st class
constraints correspond to the SGTR in 
the Lagrangian formalism.
Dirac's conjecture is rephrased in the
Lagrangian formalism as all SGTR is PETR,
and the validity of it is spoiled by
the rare example \cite{cawley_1,cawley_2} mentioned in Introduction.

\section{Canonical theory}

Canonical theory is obtained by the unique map 
from the velocity-coordinate space spanned by
$(q,u)$ to the phase space spanned by
$(q,\p)$. The map is defined by
\be
\F: (q,u) \mapsto (q,\p),
\si
\p_{A} = W_{A}(q,u)
\deff
\pdel{L}{u^A}.
\si 
({A}=1,..,n)
\eq{3_F}
\ee
$(q,\p)$'s are called canonical variables,
and let us call $\F$ velocity-
momentum map.

In a gauge theory
$\F$ is not surjection, {\it i.e.},
the image of $\F$ is not whole phase space.
Thus there exists functions $\vf_n(q,\p)$
satisfying 
\be
(q,\p)\in {\rm Im}~\F
\Longleftrightarrow
\vf_n(q,\p) =0,\sii (n=1,..,N_1)
\ee
where $N_1 = N-R$.
The condition  $\vf_n(q,\p)=0$
is called {\it primary constraints}.
Denoting the image of $\F$ by P$_1$,
the points in P$_1$ satisfy $\vf(q,\p)=0$
and for such $(q,\p)$ there exist $(q,u)$
satisfying $\p_A - W_A(q,u)=0$.

Furthermore $\F$ is not injection
in a gauge theory,
 {\it i.e.},
$W_A(q,u)=W_A(q,u')$ do not always imply 
$u=u'$.
Regarding $\p_A=W_A(q,u)$ to be defining equation of $u$,
they have solutions only if $(q,\p)\in{\rm P}_1$.
Denote the general solution of them as
\be
u^A = \hat{U}^A(q,\p),
\si (q,\p)\in {\rm P}_1.
\si (A=1,..,N)
\ee
We see
$\p_A = W_A(q,\hat{U}(q,\p))$ on P$_1$,
but $u^A \ne \hat{U}^A(q,W(q,u))$ 
in general.

We assume that $\hat{U}(q,\p)$ can be extended 
from P$_1$ to the whole phase space, preserving continuity and differentiability,
and we use the same simbol $\hat{U}^A$.
Using the extended $\hat{U}^A$ 
we can write $\vf_n$ explicily.
This is proved as follows.
If we expand lagrangian around $u=u_0=\hat{U}(q,\p)$
\be
L(q,u)
&=&
L(q,u_0)
+
(u-u_0)^{A}W_{A}(q,u_0)
+
\fr12
(u-u_0)^{A}(u-u_0)^{B}M_{AB}(q,u_0)
\nn
&& \sii\sii
+	
{\rm O}\left((u-u_0)^3\right),
\ee
then 
$\p_A - W_A(q,u)=0$ become
\be
&&
0
=
(\p_{A} - W_{A}(q,u_0))- (u-u_0)^{B}M_{AB}(q,u_0)
+
{\rm O}\left((u-u_0)^2\right).
\ee
Since the first term of r.h.s of the above equation vanishes if
$(q,\p)\in{\rm P}_1$,
the term is of order $(u-u_0)$.
Hence the necessary and sufficient condition
for the above equations to have solutions
for small $u-u_0$
is 
${\rm rank}\bbM ={\rm rank}(\bbM,\bbpi - \bbW)$.
With the regular matrix $\bbQ$ used in sweeping out $\bbM$,
the condition becomes
${\rm rank}\bbQ\bbM ={\rm rank}(\bbQ\bbM,\bbQ(\bbpi - \bbW))$.
This means the $n$th components of $\bbQ(\bbpi - \bbW)$
vanish for $R<n\le N$.
Hence the functions of the primary constraints are
found to be
\be
\vf_n(q,\p)
= 
\left[
\bbz^{(n)}\cdot(\bbpi - \bbW(q,u))
\right]_{u = \hat{U}(q,\p)},
\sii (R<n\le N)
\eq{3_def_vf}
\ee
where $\bbz$'s are the eigen vectors of the
Hessian with zero eigen value, defined by 
(\ref{1_def_z}).
Note the equations $\p_A - W_A(q,\hat{U}(q,\p)) = 0$
serve as primary constrains,
but they are not independent to each others,
while the functions defined by
(\ref{3_def_vf})
are independent because $\bbQ$ is a
regular matrix.
In what follows an equation, $R(q,\p)=0$, 
holding modulo 
primary constraints, is written as
\be
R(q,\p) = 0
\sii {\rm mod}~\vf.
\ee
For example we have
\be
\p_{A} - W_{A}(q,\hat{U}(q,\p)) 
=
0
\sii
{\rm mod}~\vf.
\eq{3_p-W=0_mod_vf}
\ee

For a function $\hat{F}(q,\p)$ on the whole
phase space the function on the velocity-coordinate space, defined by
$F(q,u)\deff \hat{F}(q,W(q,u))$
is called {\it pull-back} of $\hat{F}$,
and denote $\hat{F}_{\rm PB}$.
In what follows we will frequently use the function
defined by
\be
U^A_{\rm pb}(q,u) \deff \hat{U}^A(q,W(q,u))
=
\hat{U}^A_{\rm PB}(q,u).
\eq{2_defUpb}
\ee
Also a relation $F(q,u)=0$ derived by
$F(q,u)=\hat{F}(q,W(q,u))=0$
is called pull-back of $\hat{F}=0$.
For example the pull-back of a primary constraint
is identity.

Throughout the paper,
we use 
the following definition for the hamiltonian,
\be
H \deff
\p_A\hat{U}^A(q,\p) - L(q,\hat{U}(q,\p)),
\ee
and nothing is added.
The hamiltonian is a function on the whole phase space.
Then we see
\be
\fr{\del H}{\del q^{A}}
&=&
\left(
-
\pdel{L}{q^{A}}
+
\left.
(\p_{B} - W_{B}(q,u))
\pdel{\hat{U}^{B}}{q^{A}}
\right)
\right|_{u=\hat{U}}
=
\left.
-
\pdel{L}{q^{A}}
\right|_{u=\hat{U}}
\sii {\rm mod}~\vf,
\eq{3_dH/dq=-dL/dq}
\\ 
\pdel{H}{\p_{A}}
&=&
\hat{U}^{A}
+
\left.
(\p_{B} - W_{B}(q,u))
\pdel{\hat{U}^{B}}{\p_{A}}
\right|_{u=\hat{U}}
=
\hat{U}^{A}
\sii
{\rm mod}~\vf.
\eq{3_dH/dp=U}
\ee
An orbit O in the velocity-coordinate space 
is mapped by $\F$ to an orbit $\hat{\rm O}$ in the 
phase space.
Since $\p=W(q,\hat{U}(q,\p))$ on $\hat{\rm O}$,
by differentiating it with respect to $\t$
we have
\be
\dot{\p}_A
=
\left[
\pdel{W_A}{q^B}
\dot{q}^B
+
M_{BA}
\dot{u}^B
\right]_{u=\hat{U}}
=
-\pdel{H}{q^A}
+
\left[
{\rm [EL]_A}
+
(\dot{q}^B - u^B)
\pdel{W_A}{q^B}
\right]_{u=\hat{U}},
\eq{3_pdot=-dH/dq}
\ee
where 
we used (\ref{3_dH/dq=-dL/dq}).
Apart from the above equations,
it is necessary to get equations 
determining $\dot{q}^A$
in terms of canonical variables, for
obtaining equations of motion 
in the canonical theory.
As Dirac did almost all authors 
derive them  from variational principle
for constrained hamiltonian system.
Although the hamiltonian is defined as a function of $q,\dot{q}$ and $\p$,
its variation behaves as if it is a
function of only $q,\p$ because of the
definition of momenta.
In the case of regular system the above 
procedure is  done through the Legendre transformation which is a mathematically unambiguous tool.
But in the constrained system
this seems logically obscure.
For this reason,
Kamimura \cite{kami_1} developed a theory called
generalized canonical formalism where
the base space is spanned by
$q,\dot{q},\p$,
and introduced the concept of
generalized canonical quauntity
(GCQ) the derivatives of which with
respect to $\dot{q}$ vanish on
the primary constrained space.
In this framework, however,
the Poisson brackets among GCQ's
are not always GCQs.

In the present paper the hamiltonian 
is defined not by $\dot{q}$ but by $\hat{U}(q,p)$
which is well-defined function because
the concept of the general solution
is  mathematically sound.
Since the variational method in determining the equations for $\dot{q}$ can not be 
used here, we get the relation 
by requiring that the pull-back of
the canonical equations of motion
becomes the ELE's and 
the relation $\dot{q}=u$.
The valid choice turns out to be
\be
\dot{q}^A = \hat{U}^A(q,\p).
\eq{3_qdot=U}
\ee
Then, from (\ref{3_dH/dp=U})$\sim$
(\ref{3_qdot=U}), 
the canonical equations of motion are
\be
\dot{\p}_A = -\pdel{H}{q^A},
\sii
\dot{q}^A = \pdel{H}{\p_A}
.
\eq{3_canoEq}
\ee
In fact the pull-back of (\ref{3_canoEq})
is 
\be
{\rm [EL]}^A(U_{\rm pb}) = \dot{q}^A - U^A_{\rm pb} = 0,
\eq{3_ELEfromCEq}
\ee
where $U_{\rm pb}$ is defined by 
(\ref{2_defUpb}).
The above equations are the ELE's and the relations $\dot{q}^A - u^A$, where $u$'s are replaced by $U_{\rm pb}$'s,
and which are equivalent to the ELE's,
since the change of $u\rarw U_{\rm pb}$ is a matter of notation.

For an arbitrary function $\hat{F}(q,\p,\t)$,
(\ref{3_canoEq})
is written as
\be
\fr{d\hat{F}}{d\t} = 
\pdel{\hat{F}}{\t}
+
\{\hat{F},H\},
\ee
where the Poisson bracket is defined by
\be
\{\hat{F},\hat{G}\}
\deff
\fr{\del \hat{F}}{\del q^{A}}
\fr{\del \hat{G}}{\del \p_{A}}
-
\fr{\del \hat{F}}{\del \p_{A}}
\fr{\del \hat{G}}{\del q^{A}}
.
\ee
In what follows we use the simple notation
\be
\hat{F}^\sim \deff 
\pdel{\hat{F}}{\t}
+
\{\hat{F},H\}.
\ee
Then the canonical equation of motion is written as
$
d/d\t\hat{F}=\hat{F}^\sim
\eq{3_Fdot=Ftilde}
$.

For the complete correspondence between the
solution orbits of the lagrangian and the hamiltonian formalism, 
the solution orbit in the phase 
space should not go out of P$_1$.
This requires 
$\vf_n{}^\sim = 0$ up to the primary constraints.
There are three cases for the requirement.
(1) they are satisfied identically, (2) they restrict
the form of $\hat{U}(q,\p)$, (3) new constraints for the 
canonical variables occur.
The second cases are related to the models having  second class constraints,
and we do not consider them in the present paper.
In the third cases we write the independent relations among
$\vf_n{}^\sim=0$ as $\c_n=0$, 
and call them {\it secondary constraints}.
The time derivative of $\c_n$ must also
vanish up to the primary and the secondary
constraints, and this process continues until
there remains no conditions.
The new constraints are defined iteratively, 
in the obvious notation, as follows:
\be
\c^{(k+1)}
=
(\c^{(k)})^\sim
\si {\rm mod}~(\c^{(0)},..,\c^{(k-1)}),
\sii
\c^{(0)} = \vf,
\ee
and call $\c^{(k)}=0$ as $k$-th order 
secondary constraints.

We can show that 
the canonical equations of motion supplemented with
all of the primary and the secondary constraints
are equivalent to the ELE's 
with all of the LC's.
To see this it is sufficient to show that
the pull-black of the $k$-th order secondary constraints
are equivalent to the $k$-th order LC's.
For proving it let us calculate $(\p - W)^\sim$.
We see after rather lengthy calculations
\be
\{\p_{A} - W_A(q,\hat{U}(q,\p)),~H\}
=
\left[
-\w_A
+
(\p_B - W_B)K_A{}^B
+
M_{AB}J^B
\right]_{u=\hat{U}(q,\p)}
\eq{p-W_H}
\ee
where
\be
K_A{}^B
&\deff&
\{\p_A,\hat{U}^B\},
\nn
J^B
&\deff&
\hat{U}^C\{\p_C,\hat{U}^B\}
+
\p_D\{q^C,\hat{U}^{[D}\}\{\p_C,\hat{U}^{B]}\}
-
\{q^B,\hat{U}^C\}
\pdel{L}{q_B}
+
\{\hat{U}^B,\hat{U}^C\}
W_C.
\ee
Multiplying $z^{A(m)}(q,\hat{U}(q,\p))$,
the components of the eigenvectors of Hessian
with zero eigenvalue,
to (\ref{p-W_H}),
we see
\be
\vf_m{}^\sim
=
-
\bbz^{(m)}\cdot\bbomega\Big|_{u=\hat{U}(q,\p)}
\sii
{\rm mod}~\vf
\ee
Thus we obtain
\be
\vf_m{}^\sim\Big|_{\rm PB}(q,u)
=
-
(\bbz^{(m)}\cdot\bbomega)(q,U_{\rm pb}).
\eq{3_1stChi=1stEll}
\ee
For a functions $F(q,u)$ and $\hat{F}(q,\p)$,
if the relation 
$F(q,U_{\rm pb})=\hat{F}_{\rm PB}(q,u)$
holds, then we also call $F$ the pull-back of
$\hat{F}$ when it is not misleading. 
Then 
(\ref{3_1stChi=1stEll})
means that the pull-backs of 
1st order secondary constraints
are the 1st order LC's.

In order to extend the above result 
to higher order constraints, we use the
following relation.
That is, for functions $F(q,u,\t)$ and $\hat{F}(q,\p,\t)$
which satisfy
$F(q,U_{\rm pb},\t)=\hat{F}_{\rm PB}(q,u,\t)$, 
the following relation holds.
\be
\dot{F}(q,U_{\rm pb},\t)
=
\hat{F}^\sim
_{\rm PB}(q,\p,\t)
\sii {\rm mod}~{\rm [EM]}(U_{\rm pb}),
\eq{3_dotF=tildeF}
\ee
where ${\rm mod}~{\rm [EM]}(U_{\rm pb})$
means that the relation holds if
${\rm [EL]}(U_{\rm pb})=0$ {\it and} $\dot{q}-U_{\rm pb}=0$.
(\ref{3_dotF=tildeF})
is proved as follows.
Differentiating 
$F(q,U_{\rm pb},\t)=\hat{F}(q,W(q,u),\t)$
with respect to $\t$, we have
\be
\dot{F}(q,U_{\rm pb},\t)
&=&
\left[
\pdel{\hat{F}}{\t}
+
\dot{q}^A
\left(
\pdel{\hat{F}}{q^A}
+
\pdel{\hat{F}}{\p_B}
\pdel{W^B}{q^A}
(U_{\rm pb})
\right)
+
\pdel{\hat{F}}{\p_B}
\dot{U}_{\rm pb}^AM_{AB}(U_{\rm pb})
\right]_{\rm PB}
\nn
&=&
\left[
\pdel{\hat{F}}{\t}
+
\pdel{\hat{F}}{q^A}
U_{\rm pb}^A
+
\pdel{\hat{F}}{\p_A}
\pdel{L}{q^A}
(U_{\rm pb})
\right]_{\rm PB}
\si {\rm mod}~{\rm [EM]}(U_{\rm pb}).
\ee
On the other hand,
from
(\ref{3_dH/dq=-dL/dq}) and 
(\ref{3_dH/dp=U})
we see
\be
U_{\rm pb}^A
=
\pdel{H}{\p_A}
\Big|_{\rm PB},
\sii
-
\pdel{L}{q^A}
(U_{\rm pb})
=
\pdel{H}{q^A}
\Big|_{\rm PB}.
\eq{2_dH/dq_1}
\ee
Then we see
\be
\dot{F}(q,U_{\rm pb},\t)
&=&
\left[
\pdel{\hat{F}}{\t}
+
\pdel{\hat{F}}{q^A}
\pdel{H}{\p_A}
-
\pdel{\hat{F}}{\p_A}
\pdel{H}{q^A}
\right]_{\rm PB}
\si {\rm mod}~{\rm [EM]}(U_{\rm pb}),
\ee
which proves (\ref{3_dotF=tildeF}).

The
higher order LC's are
obtained by differentiating the 1st order one
with respect to $\t$ and using ${\rm [EM]}=0$.
Therefore from 
(\ref{3_1stChi=1stEll})
and
(\ref{3_dotF=tildeF})  
 we get the conclusion that
the pull-back of the $k$-th order secondary
constraints are equivalent to the $k$-th order LC's.

According to Dirac let us
define the concept of the 1st class
and the 2nd class constraints.
Denoting the all constraints $\f_i$, and
putting
\be
\bbX_{ij} \deff \{\f_i,\f_j\}\Big|_{\f=0},
\ee
there exists a regular matrix $\bbA$
satisfying
\be
\bbA\bbX\bbA^{-1}
=
\left(
       \begin{array}{c|c}
          \bbzero  &  \bbzero  \\ \cline{1-2}
         \bbzero &  \bbm 
       \end{array}
\right),
\sii \det{\bbm} \ne 0.
\eq{3_AXA}
\ee
Every function $\f_i$
is written as linear combination of
functions 
$\f_n^{(1)}=A_n{}^i\f_i,(n=1,..,N_{\rm c}-r)$
and
$\f_n^{(2)}=A_a{}^i\f_i,(a=r+1,..,N_{\rm c})$,
where $r$ is the rank of $\bbX$
and $N_{\rm c}$ is the number of all constraints.
The constraints 
$\f_n^{(1)}=0$
are called to belong to 1st class and
$\f_n^{(2)}=0$ to 2nd class.

Now let us consider the transformations 
in the canonical theory.
For an arbitrary function $Q$ of the
canonical variables $x=(q,\p)$, we call the transformation defined by
\be
\d_Qx = \{x,Q\}
\eq{3_def_HTR}
\ee
as {\it hamiltonian transformation},
or HTR for short.
This is the infinitesimal version of the
canonical transformations defined by
Goldstein\cite{goldstein},
especially that of the type $F_2(q,\p')$ he called.
By the definition we see the infinitesimal parameters $\hat{\e}^A=\d_Qq^A$
satisfy
\be
\pdel{\hat{\e}^A}{\p_B}
=
\pdel{\hat{\e}^B}{\p_A}
.
\eq{3_eApB=eBpA}
\ee
Putting
$
K_A\deff
\p_B\del\hat{\e}^B/\del\p_A,
$
we see
\be
\pdel{K_{A}}{\p_{B}}
=
\pdel{\hat{\e}^{B}}{\p_{A}}
+
\p_{\g}\fr{\del^2\e^{\g}}{\del\p_{B}\del\p_{A}}
.
\ee
Hence, by (\ref{3_eApB=eBpA}),
$
\del K_A/\del \p_B
$
is symmetric under exchange of $A$ and $B$,
which means there exists function
$\hat{E}(q,\p)$
satisfying
$K_A = \del \hat{E}/\del\p_A$.
That is
\be
\p_B\pdel{\hat{\e}^B}{\p_A}
=
\pdel{\hat{E}}{\p_A}
,
\eq{3_pDe/Dp=DE/Dp}
\ee
where $\hat{E}$ is determined only up to
additive function of only $q$.
Then $Q - \p_{A}\hat{\e}^{A} + \hat{E}$
is a function of only $q$'s.
In fact we see
\be
\pdel{}{\p_A}
(Q - \p_{A}\hat{\e}^{A} + \hat{E})
=\{q^A,Q\} - \hat{\e}^A = 0.
\ee
Hence we can write
\be
Q = \p_A\hat{\e}^A  - \hat{E},
\eq{3_Q=pe-E}
\ee
since the additive $q$-dependence 
can be absorbed into $\hat{E}$.

Let us consider the transformation
in the lagrangian variables defined by
\be
\d q^A = \e(q,u),
\si
\d u^A = \dot{\e}(q,u),
\si
\e(q,u) = \hat{\e}(q,W(q,u)),
\eq{3_LTRfromHTR}
\ee
which is the pull-back of the
HTR defined by
(\ref{3_def_HTR}).
The function 
$
E(q,u)\deff \hat{E}(q,W(q,u))
$
is the asosiated function
to $\e(q,u)$
in the above transformation.
In fact
\be
\pdel{E}{u^{A}}
=
\left[
\pdel{\hat{E}}{\p_{B}}
\pdel{W_{B}}{u^{A}}
\right]_{\p=W}
=
\left[
\p_{C}
\fr{\del\hat{\e}^{C}}{\del\p_{B}}
\pdel{W_{B}}{u^{A}}
\right]_{\p=W}
=
W_{C}
\pdel{\e^{C}}{u^{A}}
.
\ee
Thus we find that the pull-back of a HTR
is a LTR.
We call the lagrangian transformation
(\ref{3_LTRfromHTR})
the puff-back of the hamiltonian
transformation
(\ref{3_def_HTR}),
since
$\d_{\rm L}q^A$
is 
the pull-back of
$\d_Qq^A$.

Next let us consider
the relation between the variation of
the hamiltonian under HTR and
that of lagrangian under LTR.
We can prove
\be
\d_{Q}H
=
-\D L(q,\hat{U}(q,\p))
\sii
{\rm mod}~\vf,
\eq{3_dH_H=-dL_L}
\ee
where $\D L$ is defined by
(\ref{2_defDL}),
which is the variation
of lagrangian, dropping out total derivative
with respect to time.
We give a proof of
(\ref{3_dH_H=-dL_L})
in Appendix.

Consider the HTR generated by
a linear combination of 1st class
constraints,
\be
Q = d^m\f_m.
\ee
We call such a HTR as 
{\it Dirac transformation}, or DTR for short.
If the HTR in 
(\ref{3_dH_H=-dL_L}) is DTR, then by the construction of
the secondary constraints we see that
l.h.s of
(\ref{3_dH_H=-dL_L}), {\it i.e.}, $Q^\sim$,
is a linear combination
of the primary and the secondary
constraints.
Since the pull-back of the primary constraints
is identity and that of the secondary
constraints are LCs,
as proved before,
the pull-back of
(\ref{3_dH_H=-dL_L})
vanishes up to 
lagrangian constrains.
Thus we arrive at the first important
conclusion that the pull-back of a 
DTR is a SGTR.

The HTR,
the pull-back of which is a GTR,
is generated by $Q$ satisfying
\be
Q^\sim = 0
\sii {\rm mod}~\vf.
\ee
This is easily seen by
(\ref{3_dH_H=-dL_L}),
since
l.h.s. of
(\ref{3_dH_H=-dL_L})
vanishes up to primary constraints
and the pull-back of the equation
gives $\D L = 0$.
We call the HTR generated by such a
$Q$ as {\it canonical gauge transformation},
or CGTR for short.

Let us give  detailed 
relations among transformations
in the lagrangian and the
canonical formalism,
which are used in the next section.
First note that for a HTR generated by $Q$,
\be
\d_{\rm L}u^A \ne \d_Q\hat{U}(q,\p)\Big|_{\rm PB},
\ee
where $\d_{\rm L}$
is the LTR which is the pull-back of the
HTR.
The off-shell relation between
$\d_{\rm L}u^A$
and
$\d_Q\hat{U}^A$
is very complicated.
Fortunately we need only the on-shell
one for discussing Dirac's conjecture
in the next section,
and we can prove the following
simple relation.
If a lagrangian transformation
$\d_{\rm L}$
is the pull-back of a
DTR $\d_{Q}$,
then the following relation holds.
\be
\d_{\rm L}u^A\Big|_{u=U_{\rm pb}}
\equiv
\left[
\d_{Q}\hat{U}^A
+
\pdel{Q^\sim}{q^A}
\right]_{\rm PB}
\sii
{\rm mod}~(\ell,{\rm [EM]})(U_{\rm pb}).
\eq{3_du=dU+(q,tildeQ)}
\ee
The proof is as follows.
From
(\ref{3_dotF=tildeF})
we see
\be
\d_{\rm L}u^A\Big|_{u=U_{\rm pb}}
=
(\d_Qq^A)^\sim
\Big|_{\rm PB}
\sii
{\rm mod}~{\rm [EM]}(U_{\rm pb}).
\ee
Hence we have
\be
\d_{\rm L}u^A\Big|_{u=U_{\rm pb}}
&=&
\{\{q^A,Q\},H\}\Big|_{\rm PB}
\sii
{\rm mod}~{\rm [EM]}(U_{\rm pb})
\nn
&=&
\left[
\{\{q^A,H\},Q\} + \{q^A,\{Q,H\}\}
\right]
_{\rm PB}
\sii
{\rm mod}~{\rm [EM]}(U_{\rm pb})
\nn
&=&
\left[
\left\{\hat{U}^A + (\p_B - W_B(q,\hat{U}(q,\p))\pdel{\hat{U}^B}{\p^A}
,Q\right\}
+ 
\pdel{Q^\sim}{\p_A}
\right]
_{\rm PB}
\sii
{\rm mod}~{\rm [EM]}(U_{\rm pb}),
\nn
\eq{3_d_Lu=d_QU+}
\ee
where we used the Jacobi identity
for the equality of the second line.
The term of
$\p_B - W_B(q,\hat{U})$
is a linear combination of primary constraints.
Since $Q$ generatats a DTR,
this term vanishes except primary constraints,
 {\it i.e.},
is a linear combination of secondary constraints. 
The pull-back of it is a LC,
and we obtain 
(\ref{3_du=dU+(q,tildeQ)}).

For expressing the variation of an arbitrary function of $(q,u)$ in terms of canonical quantity
let as introduce the following brackets
\be
\{\hat{F},\hat{G}\}_{\rm M}
&\deff&
M_{AB}(q,\hat{U}(q,\p))
\pdel{\hat{F}}{\p_A}
\pdel{\hat{G}}{\p_B}
,
\\
\{\hat{F},\hat{G}\}_{\rm EM}
&\deff&
M_{AB}(q,\hat{U}(q,\p))
\pdel{\hat{F}}{\p_A}
\{\hat{U}^B,\hat{G}\}_{\rm M}.
\ee
Let us call
$\{\hat{F},\hat{G}\}_{\rm M}$
as M-bracket and
$\{\hat{F},\hat{G}\}_{\rm EM}$
as EM-bracket,
which involve informations on the
degeneracy of a gauge model
through Hessian,
and play an important role 
in the problem of Dirac's conjecture.

If $F(q,u) = \hat{F}(q,W(q,u))$,
then we can prove
\be
\d_{\rm L}F\Big|_{u=U_{\rm pb}}
=
\left[
\d_{Q}\hat{F}
+
\{\hat{F},Q^\sim\}_{\rm M}
\right]_{\rm PB}
\sii
{\rm mod}~(\ell,{\rm [EM]})(U_{\rm pb}).
\eq{3_dLF=dQF+Mbracket}
\ee
This can be proved as follows.
Varying 
$F(q,u)=\hat{F}(q,W(q,u))$
and substituting
$u$ by $U_{\rm pb}$,
we see
\be
(\d_{\rm L}F)(q,U_{\rm pb})
&=&
\left[
\pdel{F}{q^A}
\d_{\rm L}q^A
+
\pdel{F}{u^A}
\d_{\rm L}u^A
\right]_{u=U_{\rm pb}}
\nn
&=&
\left[	
\left(
\pdel{\hat{F}}{q^A}
+
\pdel{W_B}{q^A}
(U_{\rm pb})
\pdel{\hat{F}}{\p_B}
\right)
\d_{Q}q^A
+
\pdel{W_B}{u^A}
(U_{\rm pb})
\pdel{\hat{F}}{\p_B}
\left(
\d_{Q}\hat{U}^A
+
\pdel{Q^\sim}{\p_A}
\right)
\right]_{\rm PB}
\nn
&&
\sii\sii\sii\sii\sii\sii
{\rm mod}~(\ell,{\rm [EM]})(U_{\rm pb}),
\nonumber
\ee
where we used
(\ref{3_du=dU+(q,tildeQ)})
for $\d_{\rm L}u^A$.
Hence,
\be
(\d_{\rm L}F)(q,U_{\rm pb})
&=&
\left[	
\pdel{\hat{F}}{q^A}
\d_{Q}q^A
+
\pdel{\hat{F}}{\p_B}
\left(
\pdel{W_B}{q^A}
\d_{Q}q^A
+
\pdel{W_B}{u^A}
\d_{Q}\hat{U}^A
\right)
_{u=U_{\rm pb}}
\right.
\nn
&&
\sii\sii
+
\left.
M_{AB}(\hat{U}(q,\p))
\pdel{\hat{F}}{\p_B}
\pdel{Q^\sim}{\p_A}
\right]_{\rm PB}
\sii
{\rm mod}~(\ell,{\rm [EM]})(U_{\rm pb}).
\eq{3_dLF(Upb)=dQF+}
\ee
On the other hand
$\p_B=W_B(q,\hat{U}(q,\p))~$ mod $\vf$,
and 
$\d_Q\vf = \{\vf,Q\}=0~ $ mod $(\vf,\c)$
for DTR $\d_Q$.
Hence
\be
\d_Q\p_B
\Big|_{\rm PB}
&=&
\d_QW_B(q,\hat{U}(q,\p))
\Big|_{\rm PB}
\nn
&=&
\left(
\pdel{W_B}{q^A}
\d_{Q}q^A
+
\pdel{W_B}{u^A}
\d_{Q}\hat{U}^A
\right)_{u=U_{\rm pb}}
\sii {\rm mod}~(\ell,{\rm [EM]})
\ee
Substituging the above equation to
(\ref{3_dLF(Upb)=dQF+}),
we see
\be
(\d_{\rm L}F)(q,U_{\rm pb})
&=&
\left[	
\pdel{\hat{F}}{q^A}
\d_{Q}q^A
+
\pdel{\hat{F}}{\p_B}
\d_{Q}\p_B
+
M_{AB}(q,\hat{U}(q,\p))
\pdel{\hat{F}}{\p_B}
\pdel{Q^\sim}{\p_A}
\right]_{\rm PB}
\nn
&&
\sii\sii\sii\sii
{\rm mod}~(\ell,{\rm [EM]})(U_{\rm pb}),
\ee
which proves
(\ref{3_dLF=dQF+Mbracket}).

In the next section we will
encounter the functiton satisfying
\be
F(q,U_{\rm pb})=\hat{F}(q,W(q,u)),
\ee
instead of 
$
F(q,u)=\hat{F}(q,W(q,u)).
$
In this case 
the relation
(\ref{3_dLF=dQF+Mbracket}),
changes as
\be
\d_{\rm L}F\Big|_{u=U_{\rm pb}}
=
\left[
\d_{Q}\hat{F}
+
\{\hat{F},Q^\sim\}_{\rm EM}
\right]_{\rm PB}
\sii
{\rm mod}~(\ell,{\rm [EM]})(U_{\rm pb}).
\eq{3_dLF=dQF+EMbracket}
\ee
This relation can be proved in the
similar way as 
(\ref{3_dLF=dQF+Mbracket}).

As we close the section,
let us show that if a model contains 
1st class constraints and does not
2nd class ones,
then there exist CGTR.
The generator of a DTR is written as
\be
Q = \z^i\vf_i + \e^{(k)}_{i}\c_{k}^{i}.
\eq{3_existQ}
\ee
Since 2nd class constraints are absent,
there are coefficients, $c_{kij}$, satisfying
\be
\vf_{i}{}^\sim
=
c_{0ij}
\c^{j}_{1},
\sii
\c_{ki}{}^\sim
=
c_{kij}\c^{k+1}_{j},
\si (k=1,2,..).
\ee
Hence we have
\be
Q^\sim
&=&
(\e^{(1)}_i{}^\sim + c_{0ji}\z^j)
\c^j_1
+
\sum_{k=2}
(\e^{(k)}_i{}^\sim 
+ 
c_{k-1,ij}\e^{(k-1)}_{j})
\c^k_{i}
\sii
{\rm mod}~\vf.
\ee
If
\be
\e^{(1)}_i{}^\sim + c_{0ij}\z^j = 0,
\sii
\e^{(k)}_i{}^\sim 
+ 
c_{k-1,ij}\e^{(k-1)}_{j} = 0,
\eq{3_cond4Qtilde}
\ee
then $Q^\sim = 0\si{\rm mod}~\vf$.
The pull-back of (\ref{3_cond4Qtilde})
is 
\be
\dot{\e}^{(1)}_i{} + c_{0ij}\z^j = 0,
\sii
\dot{\e}^{(k)}_i 
+ 
c_{k-1,ij}\e^{(k-1)}_{j} = 0,
\si (k=2,3,..)
\eq{3_cond4Qtilde2}
\ee
which iteratively determine $\e^{(k)}_i,(k=1,2,..)$ 
except their constant modes,
if $\z^i$ are given.
Thus there exists CGTR in the model with
1st class constraints.
The pull-back of the CGTR is a GTR.

\section{Dirac's conjecture}\eq{DiracConjecture}
Before proceeding to the problem of Dirac's conjecture
let us review the Dirac theory of  
the constrained hamiltonian system.
The canonical equations of motion are obtained
by requiring that for variation of
the canonical variables the time integral
of $p\dot{q} - H_{\rm T}$ to have minimum value,
where
\be
H_{\rm T}
=
H + \l^m\vf_m,
\eq{4_def_HT}
\ee
and $\l$' s are Lagrange multipliers.
$H$ is defined to be $p\dot{q} - L$,
hence is a function of $q,\dot{q},p$.
Though $\dot{q}$ is not uniquely determined 
by the equation $p=\del L/\del {\dot{q}}$,
the ambiguity is absorbed into the 
Dirac variables explained below.
Hence $H$ is treated  as if it is 
a function of only $q,p$.

Taking into account the 1st order 
secondary constraints,
the above procedure leads to linear equations for $\l^m$,
and substituting the solutios to them back into $H_{\rm T}$
we get
\be
H_{\rm T}
=
H + v^m\vf^{[1]}_m + \l_{\rm s}^m\vf_m,
\eq{4_def_HT2}
\ee
where 
$\vf^{[1]}_m$ are 1st class primary
constraints and
$\l^m=\l_{\rm s}^m$ are special solutions to
$
\{\vf_m,\vf_n\}\l^n + \{\vf_m,H\}=0,
$
and $v^m$ are completely arbitray 
quantities.
If there are no second class constraints
then the special solutions $\l_{\rm s}^m$ are
absent, so we omit the third term in 
r.h.s. of
(\ref{4_def_HT2}).
Thus the variational problem with constraints 
is transformed to the system of
differential equations 
which are the canonical equations of
motion defined by
the modified hamiltonian
\be
H_{\rm T}
=
H + v^m\vf_m,
\eq{4_def_HT3}
\ee
where $\vf_m$ are the 1st class primary
constraints.
Finally we must include secondary constraints,
requiring
$
\{\vf_m,H_{\rm T}\}
=
\{\{\vf_m,H_{\rm T}\},H_{\rm T}\}
=
\cdots
=
0
$,
to the system of the differential
equations.

The modified hamiltonian $H_{\rm T}$
and $v^m$ are called total hamiltonian and
Dirac variables, respectively.
In many gauge theories
the original
hamiltonian $H$ contains the terms like 
$u^m\vf_m$,
where $u^m$ are unphysical variables.
This fact may be one of origins for
misunderstandings or controversies on
the problem of canonical theories.
Moreover in many gauge theories
$H$ containes the terms like $q^m\c_m$,
where $q^m$ are unphysical variables 
and $\c_m=0$ are secondary constraints.
Combining the above fact and 
the reason explained shortly,
Dirac conjectured that $\vf_m$ in
(\ref{4_def_HT3})
can be extended to all first class
constraints including secondary ones.

Now return to the problem of
Dirac's conjecture.
The solution orbit in the phase space to
the canonical equations of motion 
is uniquely determined 
if one gives Dirac variables.
Hence two states at a time are
physically equivalent if 
there are two solution orbits connecting 
the initial common state to the two states,
which are determined by two sets of
different Dirac variables.

\begin{figure}
\begc
\includegraphics[scale=0.35]{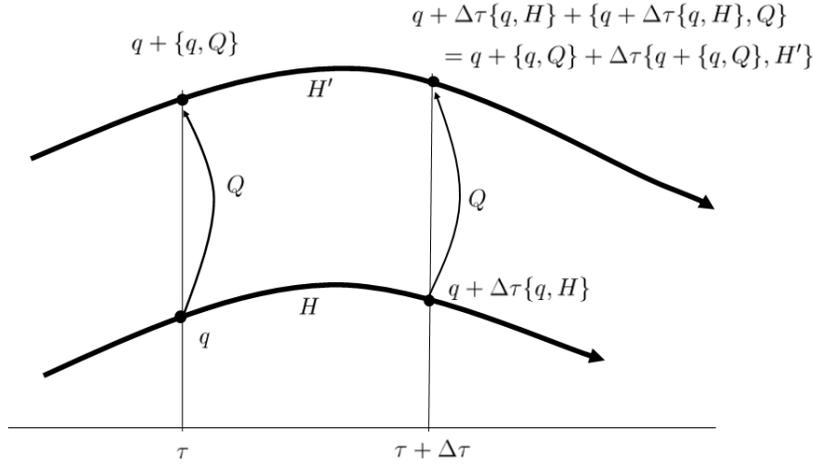}
\caption{\small Time development in phase space} 
\eq{NaiveDirac}
\endc
\end{figure}

Let us consider an solution orbit, O$_1$,
to the canonical equations of motion and its
image, O$_2$, of a transformation generated by
$Q\sim {\rm O}(\e)$.
Let $H'$ be the time development generator of
the mapped variable.
Then from the equation written at the right-top of Fig.\ref{NaiveDirac}, 
we see
$ \{q,H'-H\} = \{q,\{H,Q\}\} + {\rm O}(\e^2) $,
where the
Jacobi identity is used.
This means
$ H' = H + \{H,Q\}$.
If the transformtion is a CGTR,
the new ``hamiltonian" $H'$ is differ from
$H$ only by the Dirac variables.
Assuming the two orbits cross at some
point, they are physically equivalent.
However, the set of physically equivalent orbits
are wider than that obtained by CGTR.

A natural idea to incorporate the possible
set of physically equivalent states 
might be treating the primary and the
secondary constraints on the same footing.
Thus Dirac made the conjecture that 
a transformation generated by 
linear combination of all 1st class 
constraints including secondary ones
is a PETR.
This amounts to assume that the hamiltonian
defined by
$H_{\rm E} = H + v_i\f_i$
is the correct hamiltonian of the system,
where $\f_i$ are all 1st class constraints.

It seems hard to get the correct set of
physically equivalent class of states along
the above arguments.
Therefore, 
in the present paper, 
based on the definition of the physical
equivalence in the lagrangian theory 
and the relations of transformations between the lagrangian and the hamiltonian theories,
we seek for the condition for a DTR to be
PETR.

We assume that the initial
point does not move under the transformations
considered here.
Then the PETR is the transformation
which preserves  LC's and ELE's.
It is important to note that 
the meaning of the preservation above should be
so relaxed that the two states 
represented by $(q,u)$ and $(q',u')$
satisfying
$(q,u)\Pequiv (q',u')$ defined in
Section 2 are the same states.
Therefore we define that a LTR satisfying
\be
\d_{\rm L}\ell 
&\Pequiv&
 0 \sii{\rm mod}~(\ell,{\rm [EM]}),
\eq{4_def_PETR1}
\\
\d_{\rm L}{\rm [EL]}
&\Pequiv&
0 \sii {\rm mod}~(\ell,{\rm [EM]}),
\eq{4_def_PETR2}
\ee
is PETR.
In the phase space a HTR is defined
to be PETR if the pull-back of it is
PETR in the above sense.
For the 1st order LC's,
the condition
(\ref{4_def_PETR1})
is a special case of
(\ref{4_def_PETR2}),
while for 2nd and higher order 
LC's those conditions
are independent of
(\ref{4_def_PETR2}).

Now let us seek for the condition
that a DTR is PETR.
Since a LC is the pull-back of
a secondary constraint,
we see, from (\ref{3_dLF=dQF+Mbracket}),
\be
\d_{\rm L}\ell\Big|_{u=U_{\rm pb}}
=
\left[
\{\c,Q\} + \{\c,Q^\sim\}_{\rm M}
\right]_{\rm PB}
\sii {\rm mod}~(\ell,{\rm [EM]})(U_{\rm pb}),
\ee
where $\c$ is the secondary constraint
the pull-back of which is $\ell$.

Next let us calculate the
the variation of ELE.
From
(\ref{3_dLF=dQF+Mbracket})
we see
\be
(\d_{\rm L}W_A)(U_{\rm pb})
=
[
\d_Q\p_A + \{\p_A,Q^\sim\}_{\rm M}
]_{\rm PB}
\sii
{\rm mod}~(\ell,{\rm [EM]})(U_{\rm pb}).
\eq{3_d[EL]_1}
\ee
Hence from
(\ref{3_dotF=tildeF})
\be
\left(
\fr{d}{d\t}\d_{\rm L}W_A
\right)
(q,U_{\rm pb})
=
[
(\d_Q\p_A)^\sim + \{\p_A,Q^\sim\}^\sim_{\rm M}
]_{\rm PB}
\sii
{\rm mod}~(\ell,{\rm [EM]})(U_{\rm pb}),
\eq{3_d[EL]_1}
\ee
where 
mod $(\ell,{\rm [EM]})$
means that
the equation holds if
the quantities inside the bracket 
and their time derivatives
vanish.

From (\ref{3_dH/dq=-dL/dq})
we see
\be
\left(
\pdel{L}{q^A}
\right)(U_{\rm pb})
=
-
\left[
\pdel{H}{q^A}
\right]_{\rm PB}.
\ee
Hence from
(\ref{3_dLF=dQF+EMbracket})
\be
\left(
\d_{\rm L}
\pdel{L}{q^A}
\right)(U_{\rm pb})
=
-
\left[
\d_Q
\left(
\pdel{H}{q^A}
\right)
+
\left\{
\pdel{H}{q^A}
,Q^{\sim}
\right\}_{\rm EM}
\right]_{\rm PB}
{\rm mod}~(\ell,{\rm [EM]})(U_{\rm pb}).
\eq{3_d[EL]_2}
\ee
From
(\ref{3_d[EL]_1}) and 
(\ref{3_d[EL]_2})
we get
\be
\d_{\rm L}{\rm [EL]}_A(q,U_{\rm pb})
&=&
\left[
\fr{d}{d\t}
(\d_{\rm L}W_A)
-
\d_{\rm L}
\left(
\pdel{L}{q^A}
\right)
\right]_{u=U_{\rm pb}}
\nn
&=&
\Big[
\{\{\p_A,Q\},H\}
+
\{\{\p_A,Q^{\sim}\}_{\rm M}, H\}
\nn
&& \sii
+
\{\{H,\p_A\},Q\}
-
\{
\{\p_A,H\},Q^{\sim}
\}_{\rm EM}
\Big]
_{\rm PB}
\sii
{\rm mod}~(\ell,{\rm [EM]})(U_{\rm pb})
\nn
&=&
\Big[
\{\p_A,Q^{\sim}\}
+
\{\p_A, Q^{\sim}\}_{\rm M}
^{\sim}
-
\{\p_A{}^\sim,Q^{\sim}\}_{\rm EM}
\Big]
_{\rm PB}
\sii
{\rm mod}~(\ell,{\rm [EM]})(U_{\rm pb}),
\nn
\ee
where we used the Jacobi identity
in the second line.

The function defined by
$Q_{\x}=\x^m(q)\p_m$ with
infinitesimal parameters, 
$\x^m,~(m\ge R+1)$,
generate a PETR.
Thus we arrive at the
conclusion that
if there exist such $\x^m$
that $\hat{Q}\deff Q_\x + Q$
satisfies
\be
\{\c_i,\hat{Q}\}
+
\{
\c_i,\hat{Q}^\sim
\}_{\rm M}
= 0
\sii {\rm mod}(\vf,\c),
\eq{4_Final_1}
\ee
for all secondary constraints, $\c$'s,
and
\be
\{\p_A,\hat{Q}^{\sim}\}
+
\{
\p_A,
\hat{Q}^{\sim}
\}_{\rm M}
^{\sim}
-
\{
\p_A{}^\sim,\hat{Q}^{\sim}
\}_{\rm EM}
=
0
\si {\rm mod}~(\vf,\c),
\eq{4_Final_2}
\ee
then the DTR generated by
$Q$ is PETR.
(\ref{4_Final_1})
is the condition for the preservation of LC.
If the constraints are 
closed not only with respect to 
Poisson bracket but with respect to M-bracket,
then (\ref{4_Final_1}) is 
satisfied  with $\x^m=0$.
Let us say that this kind of constraints belonging to
class IA, and the DTR generated by them 
is of class IA DTR.
The preservation of LC is 
automatically satisfied for class IA DTR.
On the other hand
(\ref{4_Final_2})
is the condition for the preservation of ELE,
and it restricts the variation parameters 
along with the constraint structure.

Since two states which are described by
the same coordinates, $(q^A,u^A)$, except that
the $m$-components, $(q^m,u^m), (m\ge R+1)$,
are different from each other with
finite values,  
are physically equivalent,
we can extend the conditions
(\ref{4_Final_1})
and
(\ref{4_Final_2})
as
\be
\sii
\left[
\{\c_i,\hat{Q}\}
+
\{
\c_i,\hat{Q}^\sim
\}_{\rm M}
\right]_{q^m=\X^m}
=
0
\si {\rm mod}(\vf,\c),
\ee
\be
&&
\left[
\{\p_A,\hat{Q}^{\sim}\}
+
\{
\p_A,
\hat{Q}^{\sim}
\}_{\rm M}
^{\sim}
-
\{
\p_A{}^\sim,\hat{Q}^{\sim}
\}_{\rm EM}
\right]_{q^m=\X^m}
= 
0
\nn
&& \sii\sii\sii\sii\sii\sii
{\rm mod}~(\vf,\c).
\eq{4_ExtFinal}
\ee
If there exist such $\X^m$, ($m\ge R+1$), 
that at $q^m = \X^m$ the above equations hold,
then the DTR is PETR.

\section{Examples}\eq{Examples}
\subsection{Maxwell model in $D$-dimensions}

For the dynamical variables
$
q=(A^i(\bbx),A^0(\bbx)),~
u=(u^i(\bbx),u^0(\bbx))
$
the lagrangian is
\be
L
&=&
\int d^{D-1}x~{\cal L},
\nn
{\cal L}
&=&
-\fr14
F_{\m\n}F^{\m\n}
=
\fr12 m_{ij}u^iu^j
+n_iu^i
+
{\cal L}_0,
\ee
where
\be
&&
m_{ij} =\d_{ij},
\sii
n_i=\del_iA^0,
\sii
u^i = \del_0A^i,
\nn
&& \sii
{\cal L}_0
= 
-\fr12 A^0\del_i\del^iA^0
-
\fr14 F_{ij}F^{ij}
.
\ee
Here we use the metric convention:
$\h_{\m\n}={\rm diag}(-++\cdots +)$.
We use the time variable $\t = x^0$.
$W$ and the Hessian are
\be
&&
W^i
=
u^i + \del^iA^0,
\sii
W_0
=
0,
\nn
&& \si
M_{\m\n}(\bbx,\bbx')
=
\mxxb
\d_{ij}\d^{D-1}(\bbx - \bbx') & 0 \\
0 & 0
\mxxe.
\ee
ELE's and the LC are
\be
0 &=&
{\rm [EL]}_i
=
\dot{u}_i - \del_iu_0 - \del^jF_{ji},
\sii (i\le D-1)
\eq{5_MXW_eq1}
\\
0 &=&
{\rm [EL]}_0
=
\del^i(u_i - \del_iA_0)
\deff \ell,
\sii
({\rm LC})
\eq{5_MXW_eq2}
\\
0
&=&
\dot{A}^{\m} - u^{\m}.
\eq{5_MXW_eq3}
\ee
From (\ref{5_MXW_eq1}),
$\dot{\ell}$ vanishes,
hence 2nd and higher order 
LC's are absent.
The LC, (\ref{5_MXW_eq2}), is the Gauss law,
and is the Poisson equation for $A_0$,
which has the unique solution
if one sets a boundary condition for $A_0$.
However, the time derivative $u_0 = \dot{A}_0$
are arbitrary, and according to our definition, 
$A_0$ is an unphysical variable.

Under the transformation
\be
\d A_{i} = \del_i\ve,
\sii
\d A_0 = \h,
\sii
\d u_{\m} = \fr{d}{d\t}(\d A_{\m})
\eq{5_MXW_LT}
\ee
with arbitrary parameters, 
$\ve(\bbx,\t),\h(\bbx,\t)$,
the lagrangian  varies as
\be
\d L
=
\int d^{D-1}x~
(\h -\dot{\ve})\ell
+ {\rm T.D.}.
\eq{5_MXW_dL}
\ee
Hence (\ref{5_MXW_LT})
is a LTR, and also is a SGTR.

Primary constraint in the canonical theory is
\be
\vf
\deff
\p_0
=
0.
\ee
The general solution to $\p = W$
is
\be
\hat{U}^i = \p^i - \del^i A^0,
\sii
\hat{U}^0 = \th(A,\p),
\ee
where $\th$ is an arbitrary function.
Using them, hamiltonian is given by
\be
H
=
\int d^{D-1}x~
\left(
\fr12\p_i\p^i + \fr14 F_{ij}F^{ij}
+
A^0\c + \th\vf
\right),
\ee
where 
\be
\c \deff \del_i\p^i.
\ee
1st order secondary constraint
is $\c = 0$,
and 2nd and higher order
secondary constraints are absent.

The canonical equations of motion are
\be
\dot{A}^i
=
\p^i - \del^i A^0,
\sii
\dot{A^0}
=
\th(A,\p), 
\sii
\dot{\p}_i
=
\del^jF_{ji}
\sii
\dot{\p}_0
=
-\del_i\p^i.
\eq{5_MXW_Ceq}
\ee
Substituting $\p^\m = W^\m(A,u)$
into the above equations,
we get the pull-back of them as
\be
\dot{A}^i = u^i,
\sii
\dot{A}^0 = \th(A, W(A,u)),
\sii
\dot{u}_i 
=  \del_i\dot{A}_0 + \del^jF_{ji},
\sii
\ell = 0.
\eq{5_PB_CEM}
\ee
If we write
\be
U_{\rm pb} \deff \th(A, W(A,u)),
\ee
then 
(\ref{5_PB_CEM})
coincide with
the Euler-Lagrange equations,
the relations $\dot{A}^\m - u^\m=0$
and the LC, expressed in (\ref{5_MXW_eq1})
$\sim$
(\ref{5_MXW_eq3}),
where $u^0$ is replaced by $U_{\rm pb}$.

Writing a generator of
DTR as
\be
Q
=
-
\int d^{D-1}x 
~(\e\c + \h\vf),
\ee
we have
\be
&&
\d_QA_i
=
\del_i\e,
\sii
\d_QA_0
=
\h,
\\
&&
Q^\sim
=
-(\e^\sim - \h)\c
\sii {\rm mod}~\vf.
\ee
If $\h=\e^\sim$ 
the HTR generated by $Q$ is CGTR,
and the pull-back of it 
is the GTR, 
$\d_{\rm L} A_{\m}
=
\del_{\m}\e.
$

The constraints, $\vf$ and $\c$,
are of 1st class, but they do not close 
with respect to M-bracket, hence
do not belong to  class IA.
Putting $\hat{Q}= \x\p_0 + Q$,
we have
\be
\{\c,\hat{Q}\} + \{\c,\hat{Q}^\sim\}_{\rm M}
=
\D(\e^\sim - \h - \x)
\sii~~ {\rm mod}~(\vf,\c),
\ee
\be
&&
\{\p_i,\hat{Q}^\sim\} 
+ 
\{\p_i,\hat{Q}^\sim\}_{\rm M}^\sim
-
\{\p_i{}^\sim,\hat{Q}^\sim\}_{\rm EM}
=
\del_i(\e^\sim - \h - \x)^\sim
\sii {\rm mod}~(\vf,\c).
\ee
Hence 
the conditions for PETR,
expressed in (\ref{4_Final_1})-(\ref{4_Final_2}),
are satisfied by
choosing $\x=\e^\sim - \h$.
Thus we see 
every DTR in the Maxwell model
is PETR.


\subsection{Relativistic particle}

Dynamical variables 
of relativistic particle with mass $m$
in $D$-dimensional
Minkowski spacetime
are the coordinate 
$x^{\m},(\m=0,1,,..,D-1)$,
and the velocity
variables
$u^{\m},(\m=0,1,,..,D-1)$.
We use the vector notation in $D$-dimensions 
as $\bbx,\bbu$.
The lagrangian is
\be
L
=
\fr{1}{2e}
\bbu^2
-
\fr12 m^2e,
\eq{5_RP_Lag}
\ee
where $e$ is the einbein and
$\t$ parametrizes an orbit of the particle.
Including the einbein we 
write the coordinate and the velocity
variables as  
$q^A=(x^\m,e)$ 
and
$u^A=(u^\m,u)$,
respectively.

Hessian and $W$ are
\be
M_{AB}
=
\mxxb \fr{1}{e}\h_{\m\n} & 0 \\
            0            & 0
\mxxe,
\sii
W_{A}
=
\vecb
\fr{1}{e}u_{\m} \\
 0
\vece.
\ee
ELE and 1st order LC are 
\be
{\bf [EL]}
\deff
\fr{d}{d\t}
\left(
\fr{\bbu}{e}
\right)
=0,
\sii
\ell 
\deff
\fr12
\left(
\fr{\bbu^2}{e^2}
+ 
m^2
\right),
= 0.
\eq{5_RP_EM}
\ee
2nd and higher order LC's
are absent.
If $m\ne 0$
the LC can be solved
for $e$ as $e=\sqrt{-\bbu^2}/m$,
and substituting it back into the
lagrangian we get the usual action
$I = \int d\t L = -m\int \sqrt{-d\bbx^2}$.
Lagrangian of the form
(\ref{5_RP_Lag})
is useful,
since it can be used even in the case of $m=0$.
The einbein, $e$, is  unphysical because
the lagrangian does not contain $u=\dot{e}$.

Under the transformation,
\be
\d\bbx = \e\bbu,
\si
\d e = \h,
\si
\d\bbu = \fr{d}{d\t}\d\bbx,
\si
\d u = \fr{d}{d\t}\d e,
\eq{4_mRP_SGT}
\ee
with arbitrary $\e$ and $\h$,
the lagrangian varies  as
\be
\d L
=
\left(
\fr{d}{d\t}(e\e) - \h
\right)
\ell
+
\fr{d}{d\t}(\e L).
\ee
Hence the transformation is a LTR,
and also is  a SGTR.

Writing the canonical momenta of
$\bbx$ and $e$ as $\bbpi$ and $\p$,
respectively,
the primary constraint is
\be
\vf \deff \p = 0.
\ee
The general solution for $u_A$ to 
equation $\p_A=W_A(\bbx,e,\bbu,u)$
is
\be
\bbu=\hat{\bbU} \deff e\bbpi,
\sii
u = \hat{U} \deff  \th(\bbx,\bbpi,e,p),
\ee
where $\th$ is an arbitrary function.
Using them the hamiltonian is written as
\be
H = e\c + \th\p,
\sii
\c \deff \fr{1}{2}(\bbpi^2 + m^2).
\ee
1st order secondary constraint is $\c=0$,
and 2nd order and higher order
secondary constraints are absent.

The canonical equations of motion are
\be
\dot{x}^i = e\p^i,
\si
\dot{e} = \th,
\si
\dot{\p}_i = 0 \si {\rm mod}~\vf,
\si
\dot{\p} = -\c.
\ee
Substituting $\p_i = u_i/e,~\p=0$ into the above equations,
we get the pull-back of them as
\be
\fr{d}{d\t}\left(\fr{u^i}{e}\right) = 0,
\si
\c = 0,
\si
\dot{x}^i = u^i,
\si
\dot{e} = \th.
\eq{5_RP_CE}
\ee
If we write
$
u' \deff \th(q,W(q,u))
$,
then (\ref{5_RP_CE})
coincide with the ELEs, the
relations $\dot{q}^A- u^A=0$ and
the LC, where $u$'s are replaced by $u'$'s.

Writing a generator of DTR as
\be
Q
=
\ve \chi + \h\vf,
\ee
we have
\be
\d_Q\bbx = \ve \bbpi,
\sii
\d_Qe = \h,
\\
Q^\sim =
(\ve^\sim - \h)\c \si {\rm mod}~\vf.
\ee
Putting $\hat{Q} = \x\p + Q$, we see
\be
\{\c,\hat{Q}\} + \{\c,\hat{Q}^\sim\}_{\rm M} =
(\ve^\sim - \h-\x)\bbpi^2,
\sii {\rm mod}(\vf,\c)
\ee
\be
&&
\{\bbpi,\hat{Q}^\sim\}
+
\{\bbpi,\hat{Q}^\sim\}_{\rm M}
-
\{\bbpi^\sim,\hat{Q}^\sim\}_{\rm EM}^\sim
=
\left(
\fr{1}{e}(\ve^\sim - \h - \x)\bbpi
\right)^\sim
\sii
{\rm mod}(\vf,\c).
\ee
In the massless case,
the constraints, $\vf$ and $\c$ 
close with respect to M-bracket, so
belong to class IA,
while in the massive case,
they do not.
For both cases,
the conditions for PETR,
(\ref{4_Final_1}) and 
(\ref{4_Final_2}),
are satisfied by
choosing $\x=\ve^\sim - \h$.
Thus we see 
every DTR of the relativistic particle
is PETR.

\subsection{Bilocal particle}

The model consists of two relativistic particles,
and hidden symmetries extending the reparametrizations and
a global symmetry were found\cite{hori_1}.
Dynamical variables are 
$
q=(x^{\m}_k,e_k),~u=(u^{\m}_k,u_k),~(k=1,2)
$
, and lagrangian is
\be
L
=
\fr{1}{2e_1}
\bbu_1\cdot\bbu_1
+
\fr{1}{2e_2}
\bbu_2\cdot\bbu_2
+
\k(
\bbu_1\cdot\bbx_2
-
\bbu_2\cdot\bbx_1),
\eq{5_BL_L}
\ee
where $\k$
is a parameter of the model with
dimension of mass square.
ELE's and
LC's are
\be
&&
{\bf [EL]}_k
\deff
\fr{d}{d\t}
\left(
\fr{\bbu_k}{e_k}
-
2\k s_k\bbx_{k'}
\right)
=
\bbzero,
\si
({\rm ELE})
\eq{5_BL_EL}
\\
&&
\ell_k
\deff
\bbu_k\cdot\bbu_k
 = 0,
\sii (k=1,2)
\si\si
({\rm LC})
\eq{5_BL_LC}
\\
&&
\bbu_k=\dot{\bbx}_k,
\sii
u_k = \dot{e}_k,
\sii (k=1,2)
\eq{5_BL_u=dotq}
\ee
where we put
\be
s_k \deff (-1)^k,
\sii
k' \deff 3-k.
\ee
Hessian and $W$ are
\be
&&
M_{AB}
=
\mxxb \fr{1}{e_k}\d_{jk}\h_{\m\n} & 0 \\
            0            & 0
\mxxe,
\nn
&& \sii
\bbW_k
=
\fr{\bbu_k}{e_k}
-
\k s_k\bbx_{k'},
\sii
W_k = 0.
\ee
Using (\ref{5_BL_EL}),
the time derivative of $\ell_k$
are given as
\be
\dot{\ell}_k
=
4\k e_ks_k\bbu_1\cdot\bbu_2
\sii {\rm mod}~{\rm [EL]}.
\ee
Hence if 
$\k\ne 0$
we have a 
2nd order LC,
\be
\ell_0
\deff
\bbu_1\cdot\bbu_2
=
0.
\ee
3rd and higher order lagrangian constraints
are absent.
Since the lagrangian does not contain,
the variables $e_k,(k=1,2)$
are unphysical.

Under the transformation defined by
\be
&& 
\d_{\rm L} \bbx_k
=
\ve_k\bbu_k + \fr{\ve_0}{e_{k'}}\bbu_{k'},
\sii
\d_{\rm L} e_k
=\h_k,
\eq{5_BL_GTR}
\\
&&
\si
\d\bbu_k = \fr{d}{d\t}\d\bbx_k,
\sii
\d u_k = \fr{d}{d\t}\d e_{k},
\sii\sii
\ee
the lagrangian varies as
\be
\d L
=
\fr12
\left(
\fr{d}{d\t}(\ve_ke_k) - 4\k s_k\ve_0e_k
-\h_k
\right)
\ell_k
+
\fr
{
\dot{\ve}_0 - 2\k e_1e_2(\ve_1-\ve_2)
}
{e_1e_2}
\ell_0
+
{\rm T.D.}.\sii
\eq{5_BL_dL}
\ee
Hence the transformation
is LTR, and also is SGTR.

Denoting the canonical conjugates to
$\bbx_k,~e_k,(k=1,2)$ as
$\bbpi_k,~\p_k,(k=1,2)$,
the primary constraints in the canonical
theory are
\be
\vf_m \deff \p_m = 0.
\sii (m=1,2)
\ee
The general solutions of 
$
\bbu_k=\hat{\bbU}_k,
~u_k=\hat{U}_k,
~(k=1,2)
$
to the equations $\p_A = W_A(q,u)$
are
\be
\hat{\bbU}_k = e_k(\bbpi_k + \k s_k\bbx_{k'}),
\sii
\hat{U}_k = \th_k(\bbx,\bbpi,e,\p),
\ee
where $\th_k$ are arbitrary functions.
Using them the hamiltonian is written as
\be
H = e_k\c_k + \th_k\p_k,
\si
\c_k \deff \fr12
(\bbpi_k + \k s_k\bbx_{k'})^2,
\si (k=1,2).
\ee
1st order 
secondary constraints
are $\c_{1}=\c_{2}=0$,
and if $\k\ne 0$ there is a 2nd order secondary 
constraint,
\be
\c_0 \deff (\bbpi_1 - \k\bbx_2)
\cdot
(\bbpi_2 + \k\bbx_1)
=
0,
\ee
while 3rd and higher order secondary
constraints are absent.
The pull-backs of the three secondary
constraints, $\c_i=0,(i=0,1,2)$,
are equivalent to the three LCs, 
$\ell_i=0,(i=0,1,2)$.
Note that no $\c_0$ term exists in the
hamiltonian.

The Poisson brackets between $\c_i,(i=0,1,2)$
are
\be
\{\c_1,\c_2\}
=
-2\k\c_0,
\si
\{\c_k,\c_0\}
=
4\k s_k\c_k
\si (k=1,2).\si
\ee
Using the normalised basis defined by
$
\c'_1 = \c_1/2\k,~
\c'_{-1} = \c_2/2\k,~
\c'_0 = -\c_0/4\k
$, 
the Poisson brackets gives the
familiar $sl(2,\real)$ algebra:
\be
\{\c'_n,\c'_m\} = (n-m)\c'_{n+m}.
\sii (n,m=0,\pm 1)
\ee

The canonical equations of motion are
\be
&&
\dot{\bbx}_k = \hat{\bbU}_k,
\sii
\dot{e}_k =  \th_k,
\eq{5_BL_CE1}
\\
&&
\dot{\bbpi}_k
=
-
e_{k'}(\bbpi_{k'} + \k s_{k'}\bbx_k).
\eq{5_BL_CE2}
\ee
Using (\ref{5_BL_CE1}), (\ref{5_BL_CE2})
 can be written in terms of $\hat{\bbU}_k$
as
\be
\fr{d}{d\t}
\left(
\fr{\hat{\bbU}_k}{e_k}
-
2\k s_k\bbx_{k'}
\right)
= 0,
\ee
which, along with (\ref{5_BL_CE1}),
coincide with the ELE's
and  the relations $\dot{q}^A- u^A=0$,
expressed in (\ref{5_BL_EL}) and (\ref{5_BL_u=dotq}),
where $\bbu$ and $u$
are replaced by $\hat{\bbU}$ and $\th$,
respectively.

Wtiting a generator of DTR as
\be
  Q = \e_0\c_0 + \e_ke_k\c_k  + \h_k\p_k,
\ee
we have
\be
&&
\d_Q\bbx_k
=
\e_ke_k(\bbpi_k + \k s_k\bbx_{k'})
+ 
\e_0(\bbpi_{k'} - \k s_k\bbx_k), 
 \si
\d_Q\bbpi_k
=
\k s_k\d\bbx_{k'},
\si
\d_Qe_k
=
\h_k,
\\
&&
Q^{\sim}
=
A_k\c_k + A_0\c_0
\sii
{\rm mod}~\vf,
\\
&& 
A_k \deff (\e_ke_k)^\sim - 4\k s_ke_k\e_0 - \h_k,
\sii
A_0 \deff \e_0{}^\sim 
- 
2\k e_1e_2(\e_1 - \e_2).
\ee    
M-brackets among the constraints are
\be
&&
\{\vf_i,\vf_j\}_{\rm M} 
= \{\vf_i,\c_j\}_{\rm M}  
= \{\vf_i,\c_0\}_{\rm M}
= 0, 
\sii (i,j=1,2)
\\
&& 
\{\c_i,\c_j\}_{\rm M} = \fr{1}{e_i}\d_{ij}\c_i,
\sii
\{\c_0,\c_i\}_{\rm M} = \fr{1}{e_i}\c_0,
\sii
\{\c_0,\c_0\}_{\rm M} = \sum_{i=1}^2\fr{\c_i}{e_i},
\ee
hence the constraints are of class IA.
Thus the preservation of the LCs are 
satisfied automatically.
The preservation of ELE,
are written as
\be
&&
\{\bbpi,\hat{Q}^\sim\}
+
\{\bbpi,\hat{Q}^\sim\}_{\rm M}^\sim
-
\{\bbpi^\sim,\hat{Q}^\sim\}_{\rm EM}
\nn
&& \sii =
\left[
(A_k - \x_k)\fr{\hat{\bbU}_k}{e_k^2}
+
A_0\fr{\hat{\bbU}_{k'}}{e_1e_2}
\right]^\sim
=0
\si {\rm mod}~(\vf,\c)\sii~~
\ee
with $\hat{Q} = \x^m\p_m + Q$.
The first term in  the parenthesis of r.h.s. can be zero
by choosing $\x_k=A_k,(k=1,2)$,
while the second term can be zero
if, using the extended conditions for
PETR, 
\be
\e_0{}^\sim - 2\k\X_1\X_2(\e_1 - \e_2)
=0.
\ee
Since $\X_k$ are arbitrary,
the above equation can be satisfied
for arbitrary $\e_i,(i=0,1,2)$
except the case of
 $\e_1 = \e_2$.
Therefore we get the conclusion
that a DTR in the bilocal particle model is a PETR, except the 
transformation with $\e_1 = \e_2$.

\subsection{Cawley model} 

As a counter-example to Dirac's conjecture
Cawley found the following model 
\cite{cawley_1,cawley_2}.
Dynamical variables are
$(q_1,q_2,q_3)$
and the corrensponding velocity $(u_1,u_2,u_3)$,
and lagrangian is
\be
L = u_1u_2 - \fr12 q_3(q_2)^2.
\ee
ELE's and 1st order LC are
\be
&&
{\rm [EL]}_1 \deff \dot{u}_2 = 0,
\sii
{\rm [EL]}_2 \deff \dot{u}_1 + q_3q_2 = 0,
\sii
\\
&&
\ell_1 \deff q_2 = 0.
\ee
There is a 2nd order LC,
\be
\ell_2 \deff u_2 = 0,
\ee
while 3rd and higher order lagrangian
constraints are absent.
$W$ and Hessian are
\be
W_1 = u_2,
\sii
W_2 = u_1,
\sii
W_3 = 0,
\sii
M_{ij}
=
\mxxxb
  0  &  1 & 0 \\
  1  &  0 & 0 \\
  0  &  0 & 0
\mxxxe.
\ee
Since the lagrangian does not contain
$u_3$, the variable $q_3$ is unphysical.

Under the transformation
\be
\d q_1 = \e_2,
\sii
\d q_2 = 0,
\sii
\d q_3 = \h,
\eq{4_CLmodel_SGT}
\ee
lagrangian varies as
\be
\d L 
=
-\ddot{\e}_2\ell_1 + {\rm T.D.},
\ee
where O$(\ell_1)^2$ term is dropped.
Hence the transformation is
 LTR, and is also  SGTR.

Denoting the canonical momenta corresponding to
$q_i, (i=1,2,3)$ as $\p_i,(i=1,2,3)$,
the primary constraint is
\be
\vf \deff \p_3 = 0.
\ee
The general solution for $u$'s to the
equations $\p_i=W_i(q,u)$ is
\be
u_1 =\hat{U}_1 \deff \p_2, 
\si
u_2 = \hat{U}_2 \deff \p_1, 
\si 
u_3 = \hat{U}_3 \deff \th, 
\ee
where $\th$ is an arbitrary function.
Using them, hamiltonian is written as
\be
H = \p_2\c_2  + \fr12 q_3(\c_1)^2 + \th\vf,
\si
\c_1 \deff q_2,
\si
\c_2 \deff \p_1.
\ee
1st order secondary constraint is
$
\c_1 = 0.
$
There is a 2nd order secondary constraint,
$
\c_2  = 0,
$
while 3rd and higher order secondary
constraints are absent.
The pull-back of 
$\c_1$ and $\c_2$
are $\ell_1$ and $\ell_2$,
respectively.

Canonical equations of motion are
\be
&&
\dot{q}_1 = \p_2,
\si
\dot{q}_2 = \c_2,
\si
\dot{q}_3 = \th,
\\
&&
\dot{\p}_1 = 0,
\si
\dot{\p}_2 =  -q_3\c_1,
\si
\dot{\p}_3 =  -\fr12\c_1^2.
\sii
\ee
Writing the generating function $Q$ of 
DTR as
\be
Q = 
\e_1\c_1 + \e_2\c_2 + \h\vf,
\ee
we see
\be
&&
\d_Q q_1 = \e_2,
\si
\d_Q q_2 = 0,
\si 
\d_Q q_3 = \h,
\\
&&
\d_Q \p_1 = 0,
\si
\d_Q \p_2 =  -\e_1
\si
\d_Q \p_3 =  0,
\\
&&
Q^\sim =  
\e_1{}^\sim\c_1
+
(\e_2{}^\sim + \e_1)\c_2
-
\fr12\h(\c_1)^2
\si {\rm mod}~\vf.
\sii
\ee
All M-brackets among the constraints vanish, 
{\it i.e.},
they are of class IA.
Hence the preservation of the LC 
holds automatically.
On the other hand, the preservation of ELE is
written, for example, as
\be
\{\p_2,Q^\sim\}
+
\{\p_2,Q^\sim\}_{\rm M}^\sim
-
\{\p_2{}^\sim,Q^\sim\}_{\rm EM}
=
\e_2{}^{\sim\sim}
~~ {\rm mod}~(\vf,\c),
~~
\ee
which does not vanish.
R.h.s. of the above equation does not change even if we add $\x\p_3$ term to $Q$.
Thus we find that the DTR is not a PETR, 
{\rm i.e.},
Dirac's conjecture in the Cawley model 
does not hold.

\subsection{Frenkel model}

A slightly differnt model from 
Cawley's one was discussed by
Frenkel \cite{frenkel}.
The kinetic term of the lagrangian is
changed to be 3rd power of the
velocity variables, {\it i.e.},
\be
L = u_1(u_2)^2 - \fr12 q_3(q_2)^2.
\ee
ELE's are changed as
\be
{\rm [EL]}_1 \deff 2u_2\dot{u}_2 = 0,
\si
{\rm [EL]}_2 \deff 2\fr{d}{d\t}(u_1u_2) + q_3q_2 = 0.
\ee
The 1st and the 2nd order
LC's are the same as
those of Cawley model, {\rm i.e.},
$
\ell_1 \deff q_2 = 0,~
\ell_2 \deff u_2 = 0.
$
Hessian and $W$ are
\be
W_1 = (u_2)^2,
\sii
W_2 = 2u_1u_2,
\sii
W_3 = 0,
\sii
M_{ij}
=
\mxxxb
  0    &  2u_2 & 0 \\
  2u_2 &  2u_1 & 0 \\
  0    &  0    & 0
\mxxxe.
\ee

Under the transformation
\be
\d q_1 = \e_2,
\sii
\d q_2 = 0,
\sii
\d q_3 = \h,
\ee
the lagrangian varies as
\be
\d L 
=
\dot{\e}_2(\ell_2)^2
-
\fr12\h(\ell_1)^2,
\eq{5_frenkel_SGTR}
\ee
hence the transformation
is LTR, and also a SGTR.

The primary constraint is the same as
that of Cawley model:
\be
\vf \deff \p_3 = 0.
\ee
The general solution for $u$'s to the
equations $\p_i=W_i(q,u)$
is\be
u_1 = \hat{U}_1 \deff \fr{\p_2}{2\sqrt{\p_1}},
\si
u_2 = \hat{U}_2 \deff  \sqrt{\p_1},
\si
u_3 = \hat{U}_3 \deff \th.
~~
\ee
Using them hamiltonian is written as
\be
H = \p_2\sqrt{\p_1} + \fr12 q_3(q_2)^2
+
\th\vf.
\ee
1st and 2nd order secondary constraints
are the same as those of Cawley model:
\be
\c_1 \deff q_2= 0,
\sii
\c_2 \deff \p_1 = 0.
\ee
Hamiltonian is written in terms of the
constraints as
\be
H = \p_2(\c_2)^2 + \fr12 q_3(\c_1)^2 + \th\vf,
\ee
where the 1st term in r.h.s. of the
above equation is different from that
of Cawley model.
 
Writing the generating function $Q$ of 
DTR as
\be
Q = 
\e_1\c_1 + \e_2\c_2 + \h\vf,
\ee
we see
\be
&&
\d_Q q_1 = \e_2,
\sii
\d_Q q_2 = 0,
\sii 
\d_Q q_3 = \h,
\\
&&
\d_Q \p_1 = 0,
\sii
\d_Q \p_2 =  -\e_1
\sii
\d_Q \p_3 =  0,
\\
&&
Q^\sim =  
\e_1{}^\sim\c_1
+
\e_2{}^\sim\c_2
+ 
\e_1(\c_2)^2
-
\fr12\h(\c_1)^2
\si {\rm mod}~\vf.
\sii~~
\ee
The preservation of the LC's
holds automatically as in the case of Cawley
model.
The preservation of the ELE's holds also,
\be
\{\p_i,Q^\sim\}
+
\{\p_i,Q^\sim\}_{\rm M}^\sim
-
\{\p_i{}^\sim,Q^\sim\}_{\rm EM}
=
0
\sii {\rm mod}~(\vf,\c),
\ee
which is not the case in the Cawley model.
Thus we find that the DTR is a PETR,
 {\it i.e.},
Dirac's conjecture holds in the 
Frenkel model.

\subsection{Polyakov string}

Dynamical variables of the string are the coordinates, $x^\m,~(\m=0,1,..,D-1)$, of the
string in the $D$-dimensional target space.
These variables are functions of 2-dimensional
coordinates, $(\s,\t)$, and the model is 
regarded as a 2-dimensional field theory.
The velocity variables, $u^\m$, are the
$\t$-derivatives of $x^\m$. 
Lagrangian is written in the Polyakov form,
\be
L
&=&
\fr12
\int d\s~
\sqrt{g} g^{\a\b}
\del_{\a}\bbx\cdot\del_{\b}\bbx,
\ee
where $g_{\a\b}$ is the world sheet metric ($g \deff \det{g_{\a\b}}>0$), and $\bbx$ is the $D$-dimensional vector
with the components, $(\bbx)^\m=x^\m$.
Since the 2-dimensional theory
has the scale invariance,
the lagrangian is written in terms of
two variables among 
the three components of $g_{\a\b}$.
In fact, denoting 
$
a=g^{00}\sqrt{g},
~b=g^{01}\sqrt{g}
$,
the lagrangian is written as
\be
L
=
\int d\s
\left(
\fr12 a\bbu^2 + b\bbu\cdot\bbx' + \fr{1 + b^2}{2a}\bbx'^2
\right),
\eq{5_PS_Lag}
\ee
where $\bbu=\dot{\bbx}$ and $\bbx'$ are 
the derivatives 
of $\bbx$ with respect to $\t$ and $\s$,
respectively.
Since the lagrangin does not contain
the velocity variables corresponding to
$a$ and $b$,
these variables are unphysical.

ELE's and 1st order LC's are
\be
{\bf [EL]}
\deff
\fr{d}{d\t}(a\bbu + b\bbx')
+
\left(
b\bbu + \fr{1 + b^2}{a}\bbx'
\right)' = 0,
\eq{5_PS_EL}
\ee
\be
\ell_1
\deff
a^2\bbu^2 - (1 + b^2)\bbx'^2 = 0,
\si
\ell_2
\deff
a\bbu\cdot\bbx' + b\bbx'^2 = 0.
\eq{5_PS_LC}
\ee
Solving (\ref{5_PS_LC}) for $a$ and $b$,
and substituting them back into
(\ref{5_PS_Lag}),
we get the Nambu-Goto lagrangian,
$
L = \int d\s\sqrt{\det{\del_\a\bbx\cdot\del_\b\bbx}}
$.
We can check that $\dot{\ell}_i=0,(i=1,2)$
up to the ELE's and the 1st order LC's, 
so 2nd and higher order LCs are absent.
Hessian and $W$ are
\be
\bbW
=
a\bbu + b\bbx',
\sii
M_{\m\n}(\s,\s') = a\h_{\m\n}\d(\s - \s').
\ee

Under the transformation defined by
\be
\d\bbx = \e_0\bbu + \e_1\bbx',
\sii
\d a = \h_1,
\sii
\d b = \h_2,
\ee
lagrangian varies as
\be
\d L
&=& {\rm T.D.}
\sii {\rm mod}~(\ell).
\ee
Hence the transformation is LTR, and
also a SGTR.

Denoting the canonical momenta of
$\bbx$ and $(a,b)$ as
$\bbpi$ and $\p_i,(i=1,2)$,
respectively,
the primary constraints are
\be
\vf_i \deff \p_i = 0,
\sii (i=1,2).
\ee
The general solution for 
$u$'s to the equations $\p_A = W_A(q,u)$
is
\be
\bbu = \hat{\bbU} \deff 
\fr{1}{a}
(\bbpi - b\bbx'),
\si
u^i = \hat{U}^i \deff \th^i(\bbx,a,b,\bbpi,\p_1,\p_2),
\ee
where $\th_i$ are arbitrary functions.
Using them,
hamiltonian is written as
\be
&&
H = \int d\s 
\left(\fr{1}{a}\c_0
+
\fr{b}{a}\c_1
+
\th^i\vf_i
\right),
\\
&&
\c_0 \deff \fr12(\bbpi^2 - \bbx'^2),
\sii
\c_1 \deff \bbpi\cdot\bbx'.
\sii
\ee
The 1st order secondary constraints are
$\c_0=\c_1=0$,
and 2nd and higher order 
secondary constraints are absent.
The pull-back of $\c_0$ and $\c_1$ 
is equivalent to $\l_1$ and $\l_2$

Consider the DTR generated by
\be
Q = \int d\s
\left(
\e_0\c_0 + \e_1\c_1 + \h^i\vf_i
\right).
\ee
The variations of $\bbx,~a,b$,
up to the primary constraints, are
\be
\d_Q\bbx
=
\e_0\bbpi
+
\e_1\bbx',
\sii
\d_Q a_i = \h_i,
\sii (a_1=a,~a_2=b),
\ee
the pull-back of them are
the LTR with 
the redefined parameters,
$
\ve_0 = a\e_0,
~
\ve_1 = \e_1 + b\e_0
$.

The Poisson brackets among  $\c$'s and
$\vf$'s are
\be
\{\c_i(\s),\vf_j(\s')\}
=
0,
\sii\sii\sii\sii\sii\sii\sii
\ee
\be
\{\c_0(\s),\c_0(\s')\}
=
-
2\c_1(\s')\d'(\s - \s')
+
\c'_1(\s')\d(\s - \s'),
\ee
\be
\{\c_0(\s),\c_1(\s')\}
=
2\c_0(\s')\d'(\s - \s')
-
\c_0'(\s')\d(\s - \s'),
\ee
\be
\{\c_1(\s),\c_1(\s')\}
=
2\c_1(\s')\d'(\s-\s')
+
\c'_1(\s')\d(\s-\s').
\ee
Using them we have
\be
Q^\sim 
=
\int d\s
\fr{1}{a^2}(A_0\c_0 + A_1\c_1)
\sii {\rm mod}~\vf,
\ee
where
\be
A_0 
\deff
a^2\e_0{}^\sim
+
(ab\e_0 - a\e_1)'
-
2ab'\e_0
+
\h_1,
\sii\si
\ee
\be
A_1 
\deff
a^2\e_1{}^\sim
+
(ab\e_1 + a\e_0)'
-
2ab'\e_1
-
b\h_1
+
ab\h_2.
\ee

M-brackets among constraints are
\be
&&
\{\vf_i,\vf_j\}_{\rm M} 
=
\{\vf_i,\c_a\}_{\rm M} 
=
0,
\\
&&
\{\c_0(\s),\c_0(\s')\}_{\rm M} 
=
a\d(\s - \s')\bbpi^2(\s),
\\
&&
\{\c_0(\s),\c_1(\s')\}_{\rm M} 
=
a\d(\s - \s')\c_1(\s),
\\
&&
\{\c_1(\s),\c_1(\s')\}_{\rm M} 
=
a\d(\s - \s')\bbx'^2(\s).
\sii\sii\sii
\ee
Hence the constraints are not of class IA.
Putting $\hat{Q}=\x^i\p_i + Q$,
we have
\be
&&
\hat{Q}^\sim 
=
\int d\s
\fr{1}{a^2}
[
B_0\c_0 + B_1\c_1
]
\sii {\rm mod}~\vf,
\\
&&
B_0 \deff A_0 + \x^1,
\sii
B_1 \deff A_1 - b\x^1 + ab\x^2.
\sii
\ee
The condition for DTR to be PETR is,
up to $\vf$ and $\c$,
\be
&&
\{\c_0,\hat{Q}\}
+
\{\c_0,\hat{Q}^\sim\}_{\rm M}
=
B_0\bbx'^2
=
0,
\eq{5_PS_condPETR1}
\\
&&
\{\c_1,\hat{Q}\}
+
\{\c_1,\hat{Q}^\sim\}_{\rm M}
=
B_1\bbpi^2
=
0,
\sii\sii
\eq{5_PS_condPETR2}
\ee
and
\be
&&
\{\bbpi,\hat{Q}^\sim\}
+
\{\bbpi,\hat{Q}^\sim\}_{\rm M}^\sim
-
\{\bbpi^\sim,\hat{Q}^\sim\}_{\rm EM}
\nn
&&
\si
=
(B_1\bbpi - B_0\bbx')'
+
(B_0\bbpi + B_1\bbx')^\sim
\nn
&&
\sii\si
+
\left\{
\left[
\fr{1}{a}
\left(
\fr{B_0}{a^2}
\right)'
+
\fr{b}{a}
\left(
\fr{B_1}{a^2}
\right)'
\right]'
\bbpi
+
\left[
\fr{1}{a}
\left(
\fr{B_1}{a^2}
\right)'
-
\fr{b}{a}
\left(
\fr{B_0}{a^2}
\right)'
\right]'
\bbx'
\right\}'
=
0.
\sii\sii
\eq{5_PS_condPETR3}
\ee
(\ref{5_PS_condPETR1})
$\sim$
(\ref{5_PS_condPETR3}) 
are satisfied
if we choose
$\x_1 = -A_0,~\x_2 = -(A_1 + bA_2)/ab$,
 {\it i.e.}, $B_0=B_1=0$.
Thus the DTR is a PETR,
 {\it i.e.},
Dirac's conjecture holds 
in the Polyakov string .

\subsection{Model with 2nd class constraints}

In the present paper we have exclusively 
treated the gauge models which have not
2nd class constraints.
The concept of unphysical variables are
based on the absence of them.
In the final subsection
we illustrate the effect of them
in a model having such constraints.

Dynamical variables are
$x_1,x_2$ and their velocity
variables $u_1,u_2$.
Lagrangian is
\be
L
=
u_1x_2 - u_2x_1 - \fr12 (x_1)^2 - \fr12 (x_2)^2.
\ee
The ELE's do not determine the
time development of the velocity variables,
and give LC's,
\be
\ell_1 \deff 2u_2 + x_1 = 0,
\sii
\ell_2 \deff 2u_1 - x_2 = 0.
\ee
$W_i$ and Hessian are
\be
W_1 = x_2,
\sii
W_2 = -x_1,
\sii
M_{ij} = 0.
\ee
Combining the LC's and the relations $\dot{x}_i=u_i,(i=1,2)$, 
we have unique solution,
\be
x_1 = A\sin{\fr{\t}{2}} + B\cos{\fr{\t}{2}},
\sii
x_2 = A\cos{\fr{\t}{2}} - B\sin{\fr{\t}{2}},
\eq{5_2nd_sol}
\ee
with arbitrary constants, $A$ and $B$.
Hence the variables $x$'s
can not be regarded as unphysical ones,
though there are no ELE's determining 
time development of velocity variables.

The primary constraints in the canonical
theory are
\be
\vf_1 = \p_1 - x_2 = 0,
\sii
\vf_2 = \p_2 + x_1 = 0,
\ee
with the Poisson bracket,
\be
\{\vf_1,\vf_2\} = -2.
\ee
Since $W_i$ do not contain
the velocity variables,
the general solution to $\p_i=W_i$
is completely arbitrary function,
\be
u_i = U_i(x,\p).
\ee
Then the hamiltonian is
\be
H = \fr12(x_i)^2 + U_i\vf_i.
\eq{5_2nd_H}
\ee
From the preservation of $\vf_i,(i=1,2)$,
we have
\be
U_2 + \fr12 x_1 = 0,
\sii
U_1 - \fr12 x_2 = 0.
\ee 
which are the equations determining $U_i$,
and, of course, are not secondary constraints.
Substituting the above equation to
(\ref{5_2nd_H}),
the hamiltonian becomes
\be
H = \fr12(x_i)^2
+ \fr12 x_2\vf_1 - \fr12 x_1\vf_2.
\ee
According to the general prescription for
the 2nd class constraints,
the canonical equations of motion
should be
\be
\dot{x}_i = \{x_i,H\}_{\rm D},
\sii
H = \fr12(x_i)^2,
\eq{3_2nd_CEofM}
\ee
where Dirac bracket is defined by
\be
\{A,B\}_{\rm D} 
\deff 
\{A,B\} 
- \fr12\{A, \vf_1\}\{\vf_2,B\}
+ \fr12\{A, \vf_2\}\{\vf_1,B\}.
\ee
Then 
the canonical equations of motion
are
\be
\dot{x}_1 = \fr12 x_2,
\sii
\dot{x}_2 = -\fr12 x_1,
\ee
which have the solution 
expressed in (\ref{5_2nd_sol}).

Since there is no 1st class
constraints, we have no DTR nor CGTR.
Though the Hessian matrix vanishes,
there is no gauge invariance.
The reason for the absence of the
unphysical variables is the existence
of the 2nd class constraints.

 \vs{1cm}

\noi {\bf Acknowledgments}\\

The author is grateful to M.Kamata and
T.Koikawa for stimulating discussions.


\makeatletter
  \renewcommand{\theequation}{%
     A. \arabic{equation}}
  \@addtoreset{equation}{section}
\makeatother
\setcounter{equation}{0}

\section*{Appendix}

Here 
we prove the following relation.
If the LTR, 
$\d_{\rm L}$,
is the pull-back of a HTR,
$\d_Q$,
then
\be
\d_{Q}H
=
-\D L\Big|_{u=\hat{U}(q,\p)}
\sii
{\rm mod}~\vf,
\eq{A_dH_H-dL_L}
\ee
where $\D L$
is the variation of lagrangian,
defined in 
(\ref{2_defDL}),
where total derivatives with respect to time
are dropped.

In order to prove 
(\ref{A_dH_H-dL_L}) 
denote the generating function of the HTR 
in the form of (\ref{3_Q=pe-E}),
\be
Q = \hat{\e}^{A}\p_{A} - \hat{E}(q,\p).
\ee
Then the LTR is
\be
&&
\d_{\rm L}q^{A}
=
\e(q,u)
\si
\d_{\rm L}u^{A}
=
\fr{d}{d\t}(\d_{\rm L}q^{A})
\\
&&
\e(q,u) = \hat{\e}(q,W(q,u)),
\si
E(q,u) = \hat{E}(q,W(q,u)).
\sii\si
\ee
L.h.s of 
(\ref{A_dH_H-dL_L}) 
is calculated as
\be
\d_Q H
&=&
\{H,~\hat{\e}^{A}\p_{A} - \hat{E}\}
\nn
&=&
\hat{\e}^A
\{H,\p_A\}
+
\p_{A}\{H,\hat{\e}^{A}\}
- 
\{H,\hat{E}\}
\nn
&=&
-
\hat{\e}^{A}
\fr{\del L}{\del q^{A}}\Big|_{u=\hat{U}(q,\p)}
+
\p_{A}\{H,\hat{\e}^{A}\}
- \{H,\hat{E}\}
\nn
&& \sii\sii\sii
{\rm mod}~\vf,
\ee
where we used 
(\ref{3_dH/dq=-dL/dq}).
The 3rd term of r.h.s is
\be
\{H,\hat{E}\}
&=&
\fr{\del H}{\del q^{A}}
\fr{\del \hat{E}}{\del \p_{A}}
-
\fr{\del H}{\del \p_{A}}
\fr{\del \hat{E}}{\del q^{A}}
\nn
&=&
\fr{\del H}{\del q^{A}}
\p_{B}\fr{\del \hat{\e}^{B}}{\del \p_{A}}
-
\fr{\del H}{\del \p_{A}}
\left(
\d_{Q}\p_{A}
+
\p_{B}\fr{\del \hat{\e}^{B}}{\del q^{A}}
\right)
\nn
&=&
-
\fr{\del H}{\del \p_{A}}
\d_{Q}\p_{A}
+
\p_{B}
\{H,\hat{\e}^{B}\}
\nn
&=&
-
\hat{U}^{A}(q,\p)
\d_{Q}\p_{A}
+
\p_{B}
\{H,\hat{\e}^{B}\}
\sii
{\rm mod}~\vf,
\ee
where in the last line we used 
(\ref{3_dH/dp=U}).
Thus we see
\be
\d_{Q}H
=
-
\hat{\e}^{A}
\fr{\del L}{\del q^{A}}
\Big|_{u=\hat{U}}
+
\hat{U}^{A}(q,\p)
\d_{Q}\p_{A}
\sii
{\rm mod}\vf.
\eq{A_dQ_H_1}
\ee
On the other hand from
(\ref{2_defDL}))
we have
\be
&&
-
\hat{\e}^{A}\fr{\del L}{\del q^{A}}
\Big|_{u=\hat{U}}
\nn
&& \sii =
\left[
-
\D L
+
\left(
W_{A}
\fr{\del\e^{A}}{\del q^{B}}
-
\fr{\del E}{\del q^{B}}
\right)
u^{B}
\right]_{u=\hat{U}}
\nn
&& \sii =
\left[
-
\D L
+
\left(
W_{A}
\left(
\fr{\del\hat{\e}^{A}}{\del q^{B}}
-
\fr{\del\hat{\e}^{A}}{\del \p_{\g}}
\fr{\del W_{\g}}{\del q^{B}}
\right)
-
\left(
\pdel{\hat{E}}{q^{B}}
-
\fr{\del \hat{E}}{\del \p_{\g}}
\fr{\del W_{\g}}{\del q^{B}}
\right)
\right)
u^{B}
\right]_{u=\hat{U}}
\nn
&& \sii =
\left[
-
\D L
+
\left(
W_{A}
\fr{\del\hat{\e}^{A}}{\del q^{B}}
-
\fr{\del \hat{E}}{\del q^{B}}
+
(\p_{A}-W_{A})
\fr{\del\hat{\e}^{A}}{\del \p_{\g}}
\fr{\del W_{\g}}{\del q^{B}}
\right)
u^{B}
\right]_{u=\hat{U}},
\ee
where (\ref{3_pDe/Dp=DE/Dp}) is used.
Hence we see
\be
-
\hat{\e}^{A}\fr{\del L}{\del q^{A}}
\Big|_{u=\hat{U}}
=
-
\D L
\Big|_{u=\hat{U}}
-
\hat{U}^{B}(q,\p)
\d_{Q}\p_{B}
\sii
{\rm mod}\vf.
\ee
Substituting the above equation to
(\ref{A_dQ_H_1})
we get
(\ref{A_dH_H-dL_L}). \vs{1cm}


\end{document}

%% file: butsuri_article_2018.tex
\def\bfr{\begin{flushright}}
\def\efr{\end{flushright}}
\def\si{\quad}
\def\sii{\qquad}

\def\noi{\noindent}
\def\vs{\vspace}
\def\vss{\vspace{.5cm}}
\def\begc{\begin{center}}
\def\endc{\end{center}}
\def\be{\begin{eqnarray}}
\def\ee{\end{eqnarray}}
\def\eq{\label}
\def\nn{\nonumber  \\}

\def\rarwww#1{\smash{%
             \mathop{\hbox to 1.3cm{\rightarrowfill}}
             \limits^{#1}}}
\def\rarw{\rightarrow}

\def\a{\alpha}
\def\b{\beta}
\def\g{\gamma}

\def\d{\delta}
\def\D{\Delta}
\def\e{\epsilon}
\def\ve{\varepsilon}
\def\z{\zeta}
\def\h{\eta}
\def\th{\theta}

\def\k{\kappa}
\def\l{\lambda}
\def\L{\Lambda}
\def\m{\mu}
\def\n{\nu}
\def\x{\xi}
\def\X{\Xi}
\def\o{\o}
\def\p{\pi}

\def\s{\sigma}

\def\t{\tau}

\def\f{\phi}
\def\F{\Phi}
\def\vf{\varphi}
\def\c{\chi}

\def\w{\omega}

\def\fr{\frac}
\def\del{\partial}

\def\mxxb{\left( \begin{array}{cc}}           
\def\mxxe{\end{array} \right)}
\def\mxxxb{\left( \begin{array}{ccc}}         
\def\mxxxe{\end{array} \right)}
\def\mxxxxb{\left( \begin{array}{cccc}}       
\def\mxxxxe{\end{array} \right)}
\def\mxxxxxb{\left( \begin{array}{ccccc}}     
\def\mxxxxxe{\end{array} \right)}
\def\mxxxxxxb{\left( \begin{array}{cccccc}}     
\def\mxxxxxxe{\end{array} \right)}
\def\mxxxxxxxb{\left( \begin{array}{ccccccc}}     
\def\mxxxxxxxe{\end{array} \right)}
\def\kakkob{\left\{ \begin{array}{c}}
\def\kakkoe{\end{array}  \right. }
\def\vecb{\left( \begin{array}{c}}
\def\vece{\end{array} \right) }

\def\bm#1{\mbox{\boldmath #1}}

\def\real{\Bbb R}

\def\bm#1{\mbox{\boldmath #1}}
\def\bbA{\bm{$A$}}

\def\bbC{\bm{$C$}}

\def\bbM{\bm{$M$}}
\def\bbN{\bm{$N$}}

\def\bbQ{\bm{$Q$}}

\def\bbU{\bm{$U$}}

\def\bbW{\bm{$W$}}
\def\bbX{\bm{$X$}}

\def\bbm{\bm{$m$}}

\def\bbu{\bm{$u$}}

\def\bbx{\bm{$x$}}

\def\bbz{\bm{$z$}}

\def\bbomega{\bm{$\w$}}

\def\bbpi{\bm{$\p$}}
\def\bbom{\bm{$\w$}}

\def\bb1{\bm{$1$}}
\def\bbzero{\bm{$0$}}

\def\bbomega{\bm{$\w$}}

\newfont{\bg}{cmr10 scaled\magstep3}

\def\deff{\stackrel{\rm {\small def}}{=}}

\def\pdel#1#2
{
\fr{\del {#1}}{\del {#2}}
}

\def\ppdel#1#2#3
{
\fr{\del^2 {#1}}{\del {#2}\del {#3}}
}
\def\Pequiv{\stackrel{\rm P}{\sim}}

%% file: DIP_1812_01_v2.bbl
\begin{thebibliography}{99}
\bibitem{dirac_0} P.A.M. Dirac, 
Can.J.Math. {\bf 2}(1950) 129.

\bibitem{dirac_1} P.A.M. Dirac, {\it Lectures on Quantum Mechanics}, (Belfer Graduate School of Science, 1964)

\bibitem{cawley_1} Cawley,
 Phys.Rev.Lett. {\bf 42}(1979), 413.

\bibitem{cawley_2} Cawley,
 Phys.Rev. {\bf D21}(1980), 252.

\bibitem{frenkel} A.Frenkel, 
 Phys.Rev. D {\bf 21}(1982), 2986.

\bibitem{sugano_1} R.Sugano and H.Kamo, 
 Prog.Theor.Phys.{\bf 67}(1982),1966. 

\bibitem{hori_1} T. Hori, 
 J.Phys.Soc.Jpn. {\bf 61}(1992),744.

\bibitem{hori_2} T. Hori, 
 Phys.Rev. D {\bf 48}(1993), R444.

\bibitem{hori_3} T. Hori, 
 Prog.Theor.Phys.{\bf 95}(1996), 803. 

\bibitem{hori_4} T. Hori,
 Prog.Theor.Phys.{\bf 122}(2009), 323. 

\bibitem{hori_5} T. Hori,
 in {\it Advances in Quantum Theory},
 (InTech, 2012), p.51.
 
\bibitem{hori-kami} T. Hori and K. Kamimura,
Prog. Theor. Phys., {\bf 73} (1985) 476.
 
\bibitem{hori-kami-tate} T. Hori, K. Kamimura and M. Tatewaki,
Phys. Lett. {\bf B185} (1987) 367.

\bibitem{kami_1} K. Kamimura, 
 IL Nuovo Cimmento, {\bf 68B}(1982), 33.
 
\bibitem{goldstein} H. Goldstein,
 {\it Classical Mechanics}, (Addison-Wesley Pub.Comp., 1957) 
 
\end{thebibliography}
